\documentclass[aps,pre,groupedaddress,showpacs,amsmath,amssymb,preprint]{revtex4}
\usepackage{graphicx}% Include figure files
\usepackage{dcolumn}% Align table columns on decimal point
\usepackage{bm}% bold math

\begin{document}
\title{Mapping molecular models to continuum theories for partially
  miscible fluids}
\author{Colin Denniston$^{1,2}$ and Mark O. Robbins$^1$}
\affiliation{$^1$Department of Physics and Astronomy, The Johns Hopkins
  University, Baltimore, Maryland 21218, USA}
\affiliation{$^2$Department of Applied Mathematics, The University of Western 
Ontario, London, Ontario N6A 5B8, Canada}
\date{\today}

\begin{abstract}
We map molecular dynamics simulations of fluid-fluid interfaces onto
mesoscale continuum theories for partially miscible fluids.   Unlike most
previous work, we examine not only the interface order parameter and density 
profiles, but also the stress.  This allows a complete mapping from the length 
scales of molecular dynamics simulations onto a mesoscale model suitable for
a lattice Boltzmann or other mesoscale simulation method.  
Typical assumptions of mesoscale models, such as
incompressibility, are found to fail at the interface, and this has a
significant impact on the surface tension.  Spurious velocities, found in 
a number of discrete models of curved interfaces, are found to be minimized
when the parameters of the mesoscopic model are made consistent with
 molecular dynamics results.  
An improved mesoscale model is given and demonstrated to produce results 
consistent with molecular dynamics simulations for interfaces with 
widths down to near molecular size.  
\end{abstract}
\pacs{68.05.-n,47.11.+j,83.10.Gr,83.10.Mj}
%47.10.+g General theory
%47.11.+j Computational methods in fluid dynamics
%47.17.+e Mechanical properties of fluids
%47.45.Gx Slip flows
%47.55.-t Nonhomogeneous flows
%68.05.-n Liquid-liquid interfaces
%83.10.Gr Constitutive relations
%83.10.Mj Molecular dynamics, Brownian dynamics
%83.10.Rs Computer simulation of molecular and particle dynamics
%83.50.Lh Slip boundary effects (interfacial and free surface flows)
%83.50.Rp Wall slip and apparent slip
%47.60.+i Flows in ducts, channels, nozzles, and conduits 
%68.08.Bc Wetting

\maketitle

\section{Introduction}

Mesoscale continuum models are increasingly popular for 
studies of complex fluids \cite{CD98,AM98}.  Rather than using a constitutive
equation for the local stress based purely on some function of strain, these
models incorporate a dependence on the internal microstructure by including
the evolution of a local order parameter.  They have met with success in 
describing bulk properties of some materials, such as shear thinning in 
liquid crystals \cite{DOY01}, shear banding flows \cite{O00}, and phase 
ordering in binary fluids \cite{JY99}.  However, their treatment of 
interfacial stresses has not been tested for consistency with large scale 
molecular simulations.  This is an important omission since the detailed 
interfacial behavior can have a dramatic influence on macroscopic flows.

One example where molecular scale interfacial properties are important is in
pinchoff of fluid drops.  In cases where
a fluid drop breaks up due to some external force,
there can be a cascade of instabilities down
to microscopic length scales \cite{B94}.  How such instabilities are
cutoff is a question of active research \cite{ON00} with practical
applications to coatings of micro-particles \cite{CN01}. 
Another example is dynamic wetting \cite{TBR93}.  When a liquid-liquid 
interface intersects a stationary solid boundary it makes a well defined 
angle with the solid known as the contact angle.  When the solid is moving,
the {\it dynamic} contact angle $\theta_d$ is a function of the wall velocity. 
 There is significant interest in reproducing this velocity dependence 
within mesoscopic models \cite{CJ00}.   However, $\theta_d$ is affected by 
details of the fluid-fluid interface and the solid-fluid interface that are 
currently unknown.  Quantitatively reproducing these effects requires a 
detailed examination of the microscopic structure of the interfaces near 
the contact point and how this information can be mapped to the scale of 
continuum models \cite{GB99}.

The difficulty in selecting an appropriate model is that many different 
parameters in the mesoscale model influence the value of some macroscopic 
property like the surface tension.   As a 
complete set of parameters is unavailable, researchers are led to pick 
parameter sets primarily for convenience, rather than based on some 
underlying knowledge base.  This can lead to significant uncertainties as to 
the length and time scales for which these models are applicable.
Our aim in this paper is to resolve some of this uncertainty in
parameter choice, at least for the simple fluid model we examine here.
This should then serve as a guide for more complex fluids.

In this paper we examine a binary mixture of simple fluids.  
The time and distance scales involved in the hydrodynamic flow and diffusion of
such fluids are accessible to molecular dynamics simulations.  This allows
us to map out the parameters of the mesoscopic model from the molecular 
simulations.  In addition, we can test this mapping by comparing its 
predictions to simulations for a range of situations not explicitly used 
in the fits.  As methods to map parameters
from microscopic to macroscopic models are not well established, we
feel it is essential to use simple models that allow extensive testing of the
mapping.

The final mesoscale model matches changes in the local stress, as well 
as order parameter and density profiles, through interfaces in the system.  
Non-local terms in the free energy, in this case gradients of order parameter
and density, are essential for reproducing the observed microscopic stress at 
interfaces.  The results show that many common assumptions are invalid. 
For example, many models neglect density variations because the bulk fluids
are essentially incompressible.  Despite this, we find that density variations 
at the interface still have a significant effect on the interfacial tension.
Perhaps more surprising is that some of the elastic constants multiplying
gradient terms are negative:  The system is stabilized by atomic 
discreteness at short scales.

In the next section we outline our molecular and mesoscale models,
along with the methods we use to simulate them.  In Section \ref{MDmeasures} 
we describe the molecular dynamics characterization of both the bulk phases
and interfaces of the binary fluid.  Fits to various free energy functionals
are described in Section \ref{fits}.   
A comparison to some simulations of situations not used in the fitting
procedure is given in Section \ref{LBresults}.  We conclude with a
summary and discussion of implications for further work.

\section{Simulations and Models}
\label{simmod}

\subsection{Molecular Scale Model}

As we wish to explore a wide range of static and dynamic properties in our
model, it is essential that it is kept relatively simple.  We will use a
model similar to that used by a number of researchers to examine critical 
properties of fluid mixtures \cite{binder}, capillary waves \cite{grest},
and interfacial slip \cite{MRbar,DR01,DR02a}. 
The model consists of a mixture of two types of molecules, labeled $1$ and $2$.
The interactions between atoms of type $i$ and $j$ separated
by a distance $r$ are modeled using a Lennard-Jones (LJ) potential,
\begin{equation}
V_{ij}(r)=4 \epsilon_{ij} \left[ (\sigma_{ij}/r)^{12}-(\sigma_{ij}/r)^6\right],
\end{equation}
where $\epsilon_{ij}$
specifies the interaction energy and $\sigma_{ij}$ the interaction
length.  Unless specified, the force is truncated at the minimum of the 
potential $r_c=2^{1/6} \sigma_{ij}$ to improve computational efficiency.
As a result, the interactions are strictly repulsive.  For simplicity,
all molecules have the same mass $m$.  For the cases studied here we take 
$\sigma_{11}=\sigma_{22}=\sigma_{12}=\sigma$.  Two molecules of the same type 
interact with the energy scale $\epsilon_{11}=\epsilon_{22}=\epsilon$. An 
extra repulsion is added between unlike molecules 
$\epsilon_{12}=\epsilon_{21}=\epsilon(1+\epsilon^*)$ to control miscibility.  

\begin{figure}
\includegraphics[width=3.1in]{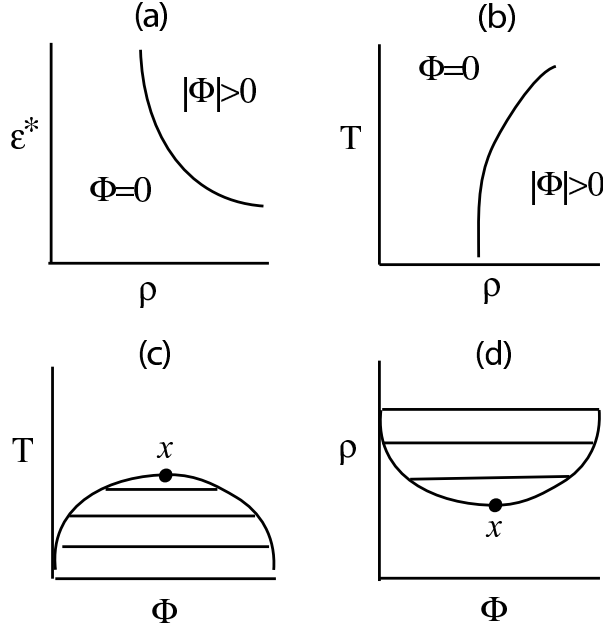}
\caption{Schematic slices through the phase diagram at (a) constant 
temperature, (b) constant $\epsilon^*$, (c) constant $\epsilon^*$ and $\rho$, 
and (d) constant $T$ and $\epsilon^*$.  In (c) and (d) the horizontal lines
tie coexisting phases and $x$ marks a critical point.}
\label{phasedia}
\end{figure}

The degree of phase separation is quantified by the order
parameter $\Phi=\rho_1-\rho_2$,  where $\rho_1$ and $\rho_2$ are the 
local densities of species $1$ and $2$.
Schematics of the phase diagram as a function of density $\rho$,
temperature $T$, and miscibility parameter $\epsilon^*$ are shown in 
Fig.~\ref{phasedia}.  For sufficiently high densities the two fluids change 
from completely miscible to completely immiscible as $\epsilon^*$ increases 
(Fig.~\ref{phasedia}(a)) or $T$ decreases (Fig.~\ref{phasedia}(b)).
For sufficiently high $\epsilon^*$ one can
observe coexistence of two partially miscible states by adjusting the
temperature or density (Fig.~\ref{phasedia}(c) and (d)).  

The object is to map the molecular model to a continuum mean field theory.  
Thus, we wish to avoid close proximity to a critical point where the behavior 
is dominated by large scale fluctuations.  However, if one is too far from the
critical point the interface may be too sharp to be described by mesoscopic
models with a continuously varying order parameter.  While not obvious a 
priori, we find that there is a significant range of parameters
where mean field theory can be used.   The constraints would be even more
relaxed in a polymer mixture where, due to the polymer
length, the interfaces are broader and the system is more mean-field like.
However, the convective-diffusive hydrodynamics of polymer molecules of any
significant length are currently inaccessible to molecular dynamics time 
scales.    

\begin{figure}
\includegraphics[width=3.1in]{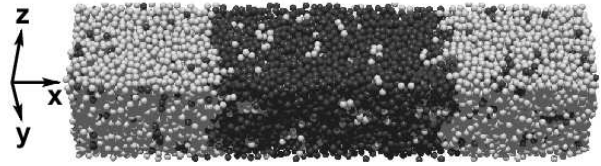}
\caption{A liquid-liquid interface between Lennard-Jones fluids at density 
  $\rho=0.85 m/\sigma^3$.
  System size is $L_x=74.36\, \sigma$ and $L_y=L_z=16.1\, \sigma$, and there
  are periodic boundary conditions in all three directions.  The
  temperature is $k_B T/\epsilon=1.1$ and $\epsilon^*=5$.}
\label{atompic}
\end{figure}

Periodic boundary conditions are applied in all three directions of the
simulation cell.  In systems with interfaces, we take the period in the 
$x$-direction $L_x$ to be $3-6$ times larger than the periods in the other 
directions, which are usually equal.  Most results are 
reported for systems with $16384$ molecules that are roughly evenly divided 
between type-1 and type-2.  The total mean density was varied between 
$\rho=0.83 \,m/\sigma^3$ and $\rho=0.925 \, m/\sigma^3$.  For the system 
with $\rho=0.85 \,m/\sigma^3$, the system size was $L_y=L_z=16.1 \sigma$ 
and $L_x=74.36 \sigma$.  A typical system configuration is shown in 
Fig.~\ref{atompic}.

To speed up the equilibration time, we supplement the molecular dynamics 
moves with a Monte Carlo procedure.  Every 500 steps, molecules are 
relabeled (from $1$ to $2$ or visa versa) according to the Metropolis 
transition rule.  This is a procedure
first used for lattice simulations \cite{binder} and later extended to
molecular dynamics simulations of polymer mixtures \cite{grest}.
After equilibration, the Monte Carlo routines are turned off. 

A characteristic time scale is given by $\tau=\sigma(m/\epsilon)^{1/2}$.
The molecular dynamics simulations use a time step $dt=0.007 \tau$.
Typically, we run the simulation for $5\times 10^5 dt$ with the Monte
Carlo routines turned on.  We wait a further $5\times 10^5 \, dt$ to ensure
that any convective flow caused by the relaxation of the initial 
(nonequilibrium) state has died away.  Data is then collected and averaged 
over the next $10^7 \,dt$.

The motion in the $y$-direction, which is always perpendicular
to any interfaces in the system, is coupled to a 
Langevin thermostat \cite{GK86, AT87}.  The drag term in the thermostat 
has a damping rate of $0.5 \tau^{-1}$.
This corresponds to under-damped motion so that even in the $y$-direction the
motion is dominated by inertia.  The equations of motion are integrated using
a velocity-Verlet algorithm \cite{AT87,lammps}.   

The local pressure tensor in a molecular dynamics simulation has the 
following form \cite{AT87}
\begin{eqnarray}
P_{\alpha\beta}({\bf r}) &=& \langle \rho v_\alpha v_\beta \rangle({\bf r}) -
  \sum_{i,j>i}\langle \frac{r_{ij\alpha}}{r_{ij}} 
  \frac{\partial V_{ij}}{\partial r_{ij}}
  \int_{C_{ij}}\hspace{-0.3cm} dr'_\beta \,\delta({\bf r}-{\bf r}')\rangle\nonumber\\
&=& \langle \rho v_\alpha v_\beta \rangle({\bf r})+
  \sum_{i,j>i} \langle f_{ij\alpha}\int_{C_{ij}}\hspace{-0.3cm} dr'_\beta
  \,\delta({\bf r}-{\bf r}')\rangle,
\label{microstress}
\end{eqnarray}
where $\langle \cdots \rangle$ denotes a thermal average, $C_{ij}$ is a 
contour joining atoms $i$ and $j$ separated by the 
vector ${\bf r}_{ij}$, and Greek indices indicate components along the
$x$, $y$, and $z$ directions.  In what follows, we adopt the summation
convention for repeated indices.
An apparent ambiguity in Eq.(\ref{microstress}) arises from the fact 
that {\it any} contour would appear to be acceptable, in that it satisfies 
microscopic momentum balance.  
However, requiring the microscopic stress to conform to the symmetry and 
transformation properties obeyed by the macroscopic stress makes the contour 
choice unique.  The appropriate contour is just the straight line between the 
two atoms, a choice originally proposed by Irving and Kirkwood
\cite{IK50}.  This is true whether or not you impose the additional
requirements on the scale of the averages in Eq. (\ref{microstress}) 
\cite{IK50,stresspaps} or on the corresponding microscopic
instantaneous many-body variable inside the averages \cite{WAD95}.

Calculating and binning the stress to get local information can be
computationally intensive as it needs to be done for every force pair
in the system.  We developed a new method to minimize this effort.
The contour line defined by ${\bf r}_{ij}$ is divided into discrete segments 
(typically four).   The contribution from each segment of the contour is then 
binned into a spatial grid every four time steps.  We choose the number of 
segments to be high enough to ensure that in steady-state systems 
the component of the pressure tensor perpendicular to fluid/fluid interfaces 
is constant, as required by conservation of momentum.

\subsection{Mesoscale Model}
\label{mesomod}

Continuum hydrodynamics is based on conservation laws and the assumption of 
local equilibrium.  The assumption of local equilibrium requires the 
existence of a free energy density.  Further, this free energy density
must be expressible as a functional of the density $\rho$, temperature $T$
and order parameter $\Phi$.  Once determined, this free energy density can 
be used to calculate the stress tensor and chemical potential required in
a full mesoscopic continuum theory. 

A number of mesoscale theories are available.  Density functional 
theories based on a weighted density approximation can reproduce
the qualitative shape of order parameter and density profiles through static 
interfaces, although the stress profile does not appear to have been 
studied \cite{TO96,IF97,NO99}.   However this approach is computationally 
challenging because it requires the use of integro-differential equations. 
Our interest is in models that are easily generalized to nonequilibrium 
situations so we restrict ourselves to the simplest ``non-local'' theory, 
namely one involving local densities and their derivatives.
 Implementations of such theories come under many names in the
literature.   Two of the most common are lattice Boltzmann simulations
\cite{SY96,CD98} and diffuse interface hydrodynamics \cite{AM98}.

To lowest order in gradients of the total density $\rho$ and the order 
parameter $\Phi$, the free energy functional can be written as
\begin{eqnarray}
{\cal F}&=& k_B T \int d{\bf r} \left\{\psi+
\frac{1}{2}  K_\rho (\nabla \rho)^2+ \right. \nonumber\\
& & \qquad \qquad \quad \left. \frac{1}{2} K_\Phi (\nabla \Phi)^2+ K_{\rho\Phi} \nabla \rho \cdot \nabla \Phi
  \right\},
\label{free}
\end{eqnarray}
where $T$ is the temperature, $k_B$ is Boltzmann's constant, and the local 
bulk free energy $\psi$ and  
elastic constants $K$ may be functions of $\rho$, $\Phi$ and $T$.
If the two phases have the same density and elastic constants (``symmetric''
case), then the free energy density must be an even function of $\Phi$.  This
requires $K_{\rho\Phi}$ to be identically zero or an odd function of $\Phi$.   
Mappings of the elastic constants used here to those of a few other 
common parameterizations of the free energy are given in Appendix \ref{Kvar}.

Conservation of the number ${\cal M}_i$ of particles of each type imposes
the constraints
\begin{eqnarray}
{\cal M}_1+{\cal M}_2 &=&\int d{\bf r}\,  \rho ,\\
{\cal M}_1-{\cal M}_2 &=& \int d{\bf r}\,  \Phi
\end{eqnarray}
Thus at equilibrium we wish to minimize the Lagrangian
\begin{eqnarray}
L={\cal F}&+&\mu_\rho k_B T\left({\cal M}_1+{\cal M}_2-\int d{\bf r}\, \rho\right)\nonumber\\
&+&\mu_\Phi k_B T\left( {\cal M}_1-{\cal M}_2-\int d{\bf r}\,  \Phi \right),
\end{eqnarray}
where $\mu_\rho$ and $\mu_\Phi$ are Lagrange multipliers, and the (constant) 
factors of $k_B T$ are included for later convenience.
The Euler-Lagrange equations give
\begin{eqnarray}
\mu_\rho&=& \frac{\partial \psi}{\partial
  \rho}-K_\rho \nabla^2 \rho-K_{\rho\Phi}\nabla^2 \Phi-\frac{\partial K_{\rho\Phi}}{\partial \Phi}(\nabla \Phi)^2,
\label{murho}\\
\mu_\Phi&=& \frac{\partial \psi}{\partial
  \Phi}-K_\Phi \nabla^2 \Phi-K_{\rho\Phi}\nabla^2 \rho.
\label{muPhi}
\end{eqnarray}
In deriving Eq.(\ref{muPhi}) we have ignored terms arising from variations
of $K_\rho$ and $K_\Phi$ with $\rho$ and $\Phi$.   Only the variation of 
$K_{\rho\Phi}$ with $\Phi$ is retained as it must be an odd function
of $\Phi$.  The equations obtained upon relaxing these assumptions
and the justification for ignoring the extra terms that result are given in
Appendix~\ref{Kvar}.

The Lagrangian density ${\cal L}$,
\begin{eqnarray}
\frac{{\cal L}}{k_B T}&=&\psi+\frac{1}{2}  K_\rho (\nabla
\rho)^2+\frac{1}{2} K_\Phi (\nabla \Phi)^2+
K_{\rho\Phi}\nabla\Phi\cdot\nabla\rho \nonumber\\
& &  -\mu_\rho \rho-\mu_\Phi \Phi,
\label{Lagrange}
\end{eqnarray}
does not contain any explicit dependence on spatial coordinates.  As a result,
Noether's Theorem \cite{G80} can be used to determine the pressure/stress
tensor ${\bf P}$.  It implies that conservation of momentum takes the form
\begin{equation}
\nabla \cdot {\bf P}=0
\end{equation}
where
\begin{equation}
P_{\alpha\beta}=-{\cal L}\delta_{\alpha\beta}+\partial_\alpha \rho
\frac{\partial {\cal L}}{\partial \partial_\beta \rho}+\partial_\alpha \Phi
\frac{\partial {\cal L}}{\partial \partial_\beta \Phi}.
\end{equation}
As this is related to conservation of momentum, we can associate 
${\bf P}$ with a nondissipative stress, or pressure tensor.
Substituting ${\cal L}$ and the Euler-Lagrange equation for $\mu_\rho$ 
gives
\begin{eqnarray}
\frac{P_{\alpha\beta}}{k_B T} &=& \frac{p_0}{k_B T}\, \delta_{\alpha\beta}
 +  K_\rho \left[(\partial_\alpha \rho)(\partial_\beta
  \rho)-\frac{1}{2}(\partial_\gamma \rho)^2
  \delta_{\alpha\beta}\right]\nonumber\\
& & + K_\Phi \left[(\partial_\alpha \Phi)(\partial_\beta
  \Phi)-\frac{1}{2}(\partial_\gamma \Phi)^2
  \delta_{\alpha\beta}\right] \nonumber\\
& & + K_{\rho\Phi}\left[(\partial_\alpha \Phi)(\partial_\beta
  \rho)+(\partial_\alpha \rho)(\partial_\beta  \Phi)\right.\nonumber\\
& &\left. \qquad\qquad -(\partial_\gamma \Phi)(\partial_\gamma \rho)\delta_{\alpha\beta}\right],
\label{PressureTensor}
\end{eqnarray}
where the hydrostatic pressure $p_0$ (trace of $\frac{1}{3}P_{\alpha\beta}$) is
\begin{equation}
p_0=k_B T(\rho\mu_\rho+\Phi \mu_\Phi-\psi),
\label{hystatpress}
\end{equation}
and the expressions for $\mu_\Phi$ and $\mu_\rho$ are given in 
Eqs. (\ref{murho}) and (\ref{muPhi}).
The relations for the pressure tensor $P_{\alpha\beta}$ and the chemical
potential $\mu_\Phi$ will allow a direct comparison between the molecular
and mesoscopic scale.

\subsection{Lattice Boltzmann Algorithm}

Mesoscale schemes, of which lattice Boltzmann algorithms are an example, 
have proved very successful in simulations of complex fluids \cite{CD98}.  
Lattice Boltzmann algorithms can be viewed as a slightly unusual 
discretization of the equations of motion or as a lattice version of a 
simplified Boltzmann equation.  In the free energy implementation, 
equilibrium properties of the fluid are naturally incorporated into the 
algorithm using ideas from statistical mechanics \cite{SY96}.  We use a 
nine velocity model on a square lattice.  Minor problems related to 
Galilean invariance \cite{SY96} are removed via a correction term in 
the second moment \cite{HB98}.

The lattice Boltzmann scheme simulates the full dynamical equations of
convective-diffusive hydrodynamics.  These include the continuity equation
\begin{equation}
\partial_t \rho+\partial_\alpha \rho u_\alpha=0,
\end{equation}
where ${\bf u}$ is the fluid velocity and Greek indices indicate directions.
Conservation of momentum takes the form
\begin{eqnarray}
& &\partial_t \rho u_\alpha+\partial_\beta \rho u_\alpha u_\beta=
-\partial_\beta P_{\alpha\beta}\nonumber\\
& &\quad +\frac{\rho \tau_f}{3}\partial_\beta\left[
(1-3 \frac{\partial p_0}{\partial \rho})
\delta_{\alpha\beta}\partial_\gamma u_\gamma+
\partial_\alpha u_\beta+\partial_\beta u_\alpha \right],
\end{eqnarray}
where the parameter $\tau_f$ sets the viscosity $\eta$.
Finally, there is the convection-diffusion equation,
\begin{equation}
\partial_t \Phi+\partial_\alpha \Phi u_\alpha= \tau_g\left(
\Gamma \nabla^2 \mu_\Phi -\partial_\beta\left(
\frac{\Phi}{\rho}\partial_\alpha P_{\alpha\beta}\right)\right).
\end{equation}
The second term on the right, describing a flow-induced diffusion \cite{J97}, 
is a common feature of lattice Boltzmann schemes and is usually negligibly 
small.  We set the algorithmic time constant $\tau_g$ to unity so that 
$\Gamma$ is the diffusion constant.  Viscous stresses associated with velocity 
gradients are not relevant here as all velocities are zero. However, 
we will make use of these terms in future works.

To compare the results of the lattice Boltzmann simulations to
molecular dynamics, it is useful to compare the units used in the
simulations.  In molecular dynamics we use Lennard-Jones units,
that is $m$, $\sigma$, and $\tau$.  Time is discrete with steps of
$dt=0.007 \tau$.  The size of the time step is limited by the requirement 
that molecular motion on the steep repulsive part of the intermolecular 
potential must be resolved.  In the lattice Boltzmann simulations 
a discretization of both space $\Delta x$ and time $\Delta t$ is required.  
We will choose $\Delta x \approx \sigma$ primarily based on the requirement 
of resolving the interface 
width.  One could, in principle, use a much coarser lattice in the bulk 
regions.  Stability of our lattice Boltzmann scheme requires the lattice 
Mach number to be less than one.  That is, in units where 
$\Delta x \equiv \Delta t \equiv 1$ , the speed of sound
\begin{equation}
v_s\approx \sqrt{\partial p_0/\partial \rho} < 1.
\end{equation}
In Lennard-Jones units $v_s \approx \sqrt{50} \sigma/\tau$ for 
$\epsilon^*=5$ (see below) so we take $\Delta t=0.1 \tau$ so that 
$v_s \approx 0.7 \Delta x/\Delta t$.

As discussed in Section~\ref{fits}, the parameters determined from molecular 
simulations turn out to be qualitatively different from those commonly used 
in lattice Boltzmann simulations.  
As such, we had some problems with numerical stability using standard schemes. 
Stability was improved by using the predictor-corrector scheme \cite{DOY01b}, 
rather than the standard Euler scheme.  Stability can be further enhanced by 
iterating the corrector step a few times.  This was found to be helpful in 
the initial steps, especially if a particularly poor initial state was used.  
In addition, the method for discretization of derivative operators, 
particularly Laplacian operators, made a significant difference.  Including a 
mixture of derivatives along coordinate directions and those taken along the 
diagonal direction improved stability.

To fully specify the model for the lattice Boltzmann algorithm, in addition to 
Eq.(\ref{PressureTensor}) for the pressure tensor, we need explicit expressions
for $\mu_\Phi$ and $p_0$.  These will be derived and fit in later sections but
for reference we list the complete expressions here:
\begin{eqnarray}
\mu_\Phi &=& 
\frac{2 A_2}{Z}\left[ e_l (\Phi+\Phi_{co})+e_r (\Phi-\Phi_{co})-2 e_r e_l \Phi \right] \nonumber\\
& & -K_\Phi \nabla^2 \Phi-K_{\rho\Phi}\nabla^2 \rho,\\
%\end{eqnarray}
%\begin{eqnarray}
p_0 &=& \rho \frac{\partial A_0}{\partial \rho}-A_0 +
\frac{1}{Z}\left(\rho\frac{dA_2}{d\rho}-A_2\right)\times\nonumber\\
& & \left[e_l (\Phi+\Phi_{co})^2+e_r (\Phi-\Phi_{co})^2-2 e_r e_l (\Phi^2+\Phi_{co}^2)\right] \nonumber\\
& & +\frac{2 \rho A_2}{Z}\frac{d\Phi_{co}}{d\rho}\left[
e_l (\Phi+\Phi_{co})-e_r (\Phi-\Phi_{co})-2 e_r e_l \Phi\right] \nonumber\\
& & -\rho\left(K_\rho \nabla^2 \rho + K_{\rho\Phi} \nabla^2 \Phi+\frac{\partial K_{\rho\Phi}}{\partial \Phi} (\nabla \Phi)^2 \right),
\end{eqnarray}
where
\begin{eqnarray}
e_r &=& \exp(-A_2(\Phi-\Phi_{co})^2/(2\rho k_B T)),\nonumber\\
e_l &=& \exp(-A_2(\Phi+\Phi_{co})^2/(2\rho k_B T)),\nonumber\\
Z &=& e_r + e_l - e_r e_l.
\end{eqnarray}

\section{Molecular Dynamics Measurements}
\label{MDmeasures}

As seen in Fig.~\ref{phasedia}, there are a number of parameters
which affect the phase (mixed or separated) of the system.   Much
of the interesting behavior in binary fluids occurs when two phases
coexist, so we will work with systems near the coexistence line.
Coexisting phases have the same $T$ and $\epsilon^*$, but $\rho$ and
$\Phi$ vary through the interface.  Unless otherwise stated, simulations 
are performed at $k_B T/\epsilon=1.1$.  The free energy functional 
Eq.~\ref{free} is then determined from simulations at $\epsilon^*=5$ and $6$.

\begin{figure}
\includegraphics[width=3.2in]{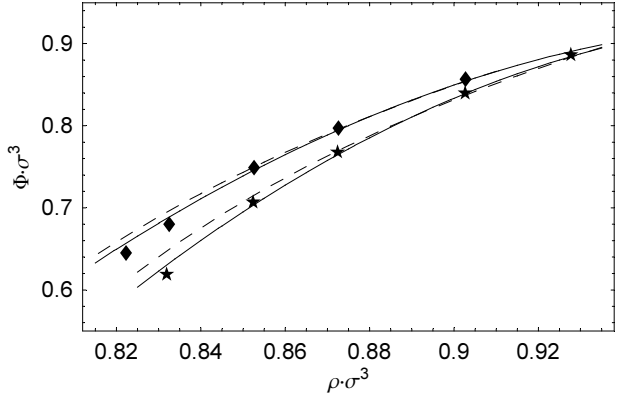}
\caption{Portion of the coexistence curve for $\epsilon^*=5$($\star$) and 
$\epsilon^*=6$ ($\blacklozenge$). The solid line is a quadratic fit to 
the data and the dashed line is from a fit to Flory-Huggins theory described
in Appendix \ref{psivar}.  Error bars are comparable to the symbol size.}
\label{rhophi}
\end{figure}

The location of the coexistence line $\Phi_{co}(\rho)$, a section of
which is shown in Fig.~\ref{rhophi}, can be easily found
by varying the average density and letting the system equilibrate (a
process we accelerate using the Monte Carlo routines).  To obtain further
information about the free energy in the vicinity of an equilibrium state,
we first study the linear response of the system to small perturbations.  
As we show in Section~\ref{linresponse}, this furnishes us with the second 
derivatives of the bulk free energy $\psi$ and the elastic constants in the 
equilibrium state.  Section~\ref{MDinterface} considers interfacial behavior.
The surface tension is evaluated as a function of $\rho$ and $T$, and 
the effect of capillary waves on the measured interface shape is 
examined.

\subsection{Linear Response of Equilibrium States}
\label{linresponse}

We begin with an equilibrated single phase configuration from a 
molecular dynamics simulation.  Thus the 'basis' state has no gradients 
of any kind.  We then add a small force on atoms of type i,
$F_i =\delta_i \cos(q x)$.  The forces can be incorporated into the chemical 
potentials, Eq.(\ref{muPhi}), as
\begin{eqnarray}
\mu_\rho^F&=& \mu_\rho -\frac{\delta_1 + \delta_2}{2q k_B T} \sin(q x)+ {\cal O}(\delta^2),\nonumber\\
\mu_\Phi^F&=& \mu_\Phi -\frac{\delta_1 - \delta_2}{2q k_B T} \sin(q x)+ {\cal O}(\delta^2),
\end{eqnarray} 
where $\mu_\rho$ and $\mu_\Phi$ are given in Eq.(\ref{muPhi}).
If the $\delta_i$ are small, we expect a linear response.  Then 
\begin{eqnarray}
\rho&=&\rho_0+\rho_\delta \sin(q x)+{\cal O}(\delta^2),\nonumber\\
\Phi&=&\Phi_0+\Phi_\delta \sin(q x)+{\cal O}(\delta^2),
\end{eqnarray}
where $\rho_0$ and $\Phi_0$ represent the undisturbed state.
Plugging this into the previous equations and expanding about the 
equilibrium state gives
%\begin{widetext}
\begin{equation}
\left( \begin{matrix} L_{\rho\rho} &  L_{\rho\Phi} \\
      L_{\Phi\rho} & L_{\Phi\Phi} \end{matrix}\right)
\left( \begin{matrix} \rho_\delta \\ \Phi_\delta \end{matrix} \right)\equiv
\left( \begin{matrix} \frac{\partial^2 \psi}{\partial \rho^2}+K_\rho q^2 &  
      \frac{\partial^2 \psi}{\partial \rho \partial \Phi}+K_{\rho\Phi} q^2 \\
      \frac{\partial^2 \psi}{\partial \rho \partial \Phi}+K_{\rho\Phi} q^2 &
      \frac{\partial^2 \psi}{\partial \Phi^2}+K_\Phi q^2 \end{matrix}\right)
\left( \begin{matrix} \rho_\delta \\ \Phi_\delta \end{matrix} \right)=
\left( \begin{matrix} \frac{\delta_1 + \delta_2}{2q k_B T} \\ \frac{\delta_1 - \delta_2}{2q k_B T} \end{matrix} \right)+{\cal O}(\delta^2),
\label{linreseqn}
\end{equation}
%\end{widetext}
where the derivatives are about the equilibrium state.
By varying the $\delta_i$ and wave vectors $q$ we can determine the second 
derivatives of $\psi$ and the elastic constants $K$ of the equilibrium state.

\begin{figure}
\includegraphics[width=3.1in]{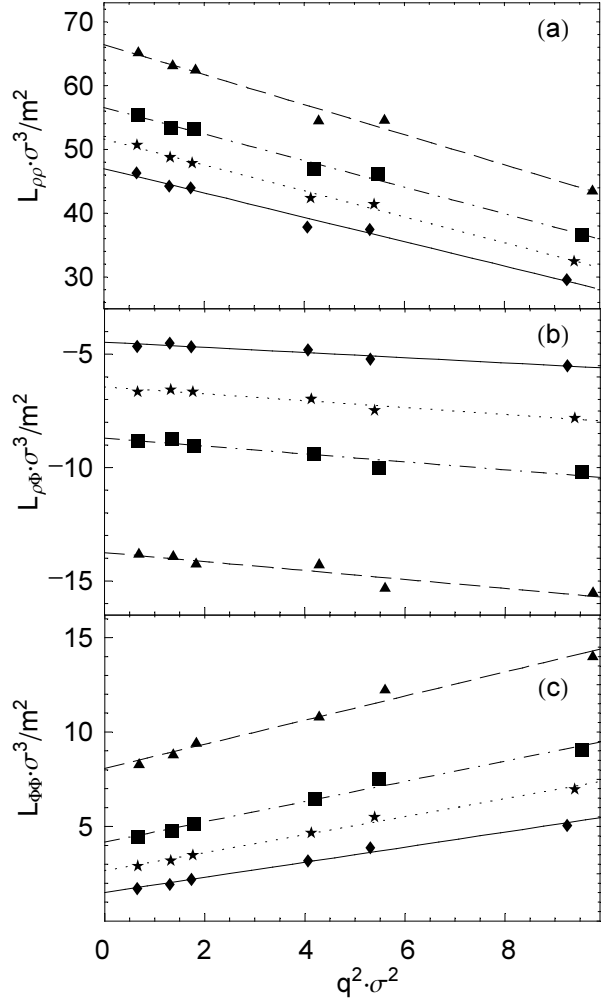}
\caption{Linear response coefficients as a function of wave vector squared.
  The slope of each linear fit (lines) gives an elastic constant and the 
  intercept gives a second derivative of $\psi$.  The temperature is 
  $k_B T/\epsilon=1.1$ and $\epsilon^*=6$. 
  Average density is 0.83 ($\blacklozenge$), 0.85 ($\star$), 
  0.87 ($\blacksquare$), or 0.9 $m/\sigma^3$ ($\blacktriangle$).  
  Statistical error bars are comparable to symbol sizes.}
\label{Lmat}
\end{figure}

To separate out the $q$ dependence we use the following technique.  
For a given $q$ we run two simulations, one with $\delta_1=\delta_2=0.2$ and 
another with $\delta_1=-\delta_2=0.1$.
This gives us four equations to determine the four elements of the 
coefficient matrix (in reality, we have one extra equation due to the 
symmetry of the matrix).  This process must be repeated for a number of 
different $q$ in order to separate out the $q$ dependence of each term.  
We obtain results for six $q$'s simultaneously by apply perturbations at 
two different (and incommensurate) q's in each of the $x$, $y$, and $z$ 
directions.

The results for ${\bf L}$ can be seen in Fig.~\ref{Lmat}.  The first thing 
to note is that the matrix elements are indeed linear functions of $q^2$.  This
indicates that the square gradient theory, assumed in Eq.(\ref{muPhi}), is 
capable of describing this system even at wavelengths as short as $2\sigma$.  
A surprising result is that both $L_{\rho\rho}$ and $L_{\rho\Phi}$
have a negative slope, implying that $K_\rho$ and $K_{\rho\Phi}$
are both {\it negative}.  This means that the square gradient theory
becomes unstable to fluctuations at short wavelengths 
($\sim 1 \sigma$) where one of the eigenvalues of ${\bf L}$ crosses zero. 
As this distance is comparable to the molecular separation,  
fluctuations on shorter scales are unphysical.  However, this sets a hard 
lower limit for the mesh spacing introduced in the lattice Boltzmann 
simulations.  Using a mesh spacing shorter than the typical molecular 
separation to ``improve accuracy'' is not only pointless, but will be 
unstable \cite{footnote}. 

It is also interesting to note that $L_{\rho\rho}$ 
is considerably larger ($5-10$ times) than the other components.  
This reflects the fact that it is harder to create fluctuations 
in density than fluctuations in concentrations.  Conversely,
a small change in density, such as the one seen in Fig.~\ref{Phirho} at an 
interface, may cost a non-negligible amount of free energy and thereby 
contribute significantly to the surface tension.  We will discuss this
point further in the next section.

\begin{figure}
\includegraphics[width=3.1in]{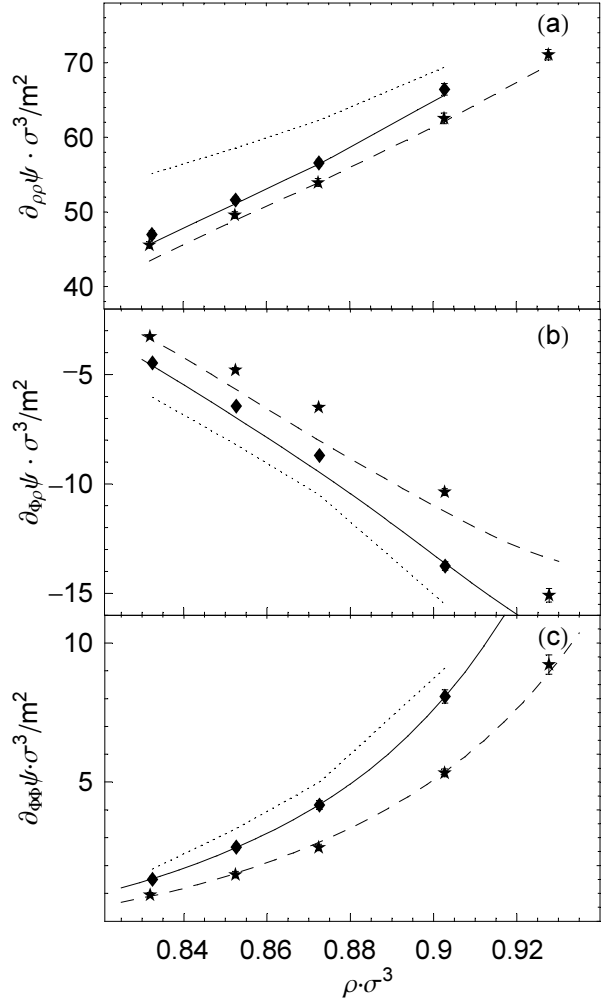}
\caption{The second derivatives of $\psi$ as a function of 
density along the coexistence line for ($\blacklozenge$) $\epsilon^*=6$ 
and ($\star$) $\epsilon^*=5$.  Data points were obtained from the intercepts
of the linear fits in Fig.~\ref{Lmat}.   
The second derivatives of the parameterization of $\psi$ for $\epsilon^*=6$ 
(solid line) and $\epsilon^*=5$ (dashed line) given in
Section~\ref{fits} and the 
Flory-Huggins free energy (dotted line) discussed in Appendix~\ref{psivar} 
for $\epsilon^*=6$ are also shown.}
\label{d2fit}
\end{figure}

\begin{figure*}
\includegraphics[width=6.4in]{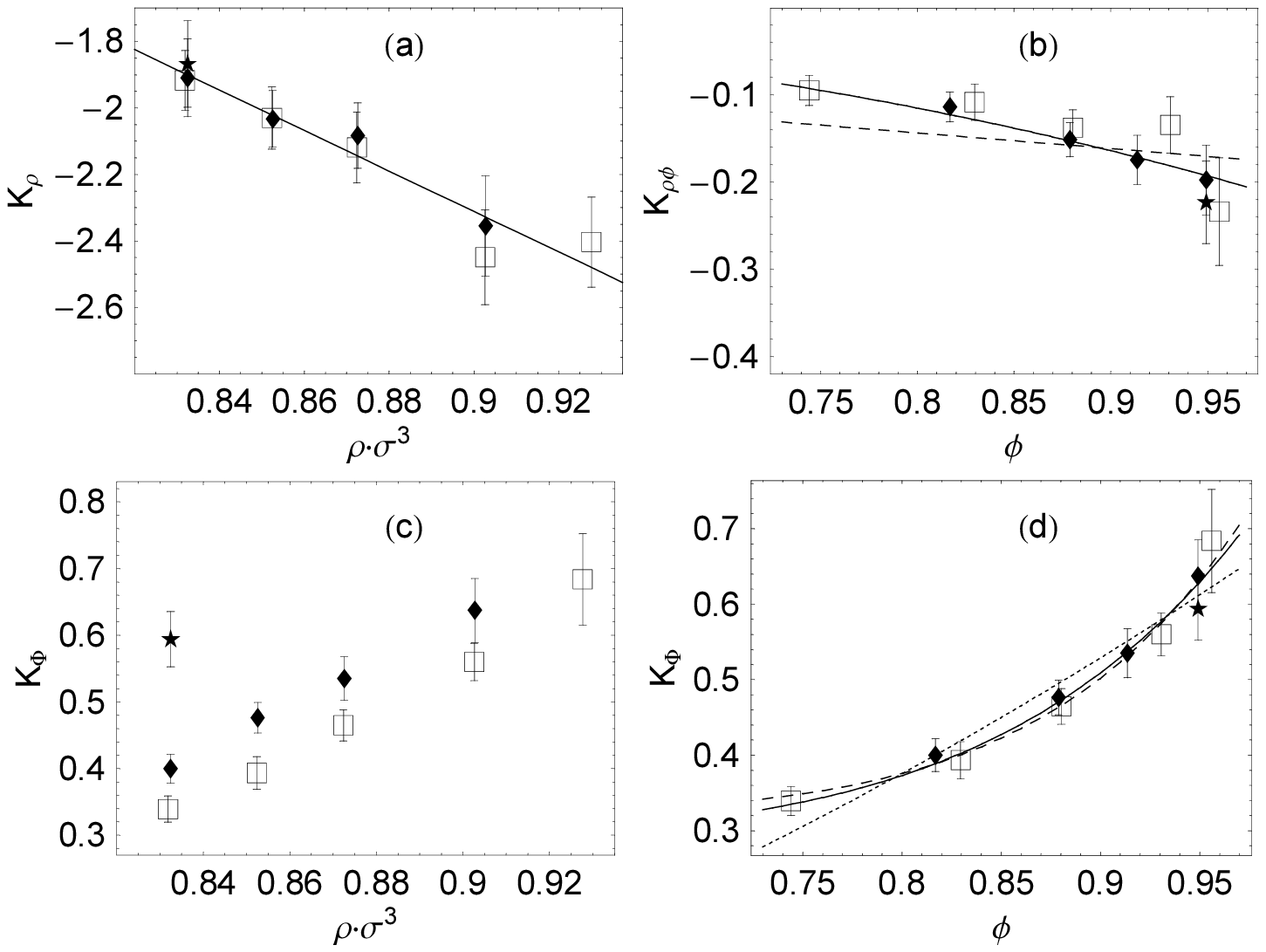}
\caption{The  elastic constants $K_\rho$, $K_\Phi$, and 
$K_{\rho\Phi}$ as a function of the density $\rho$ or order parameter
$\phi=\Phi/\rho$.  The points with the $\Box$ symbol are for $\epsilon^*=5$
and the $\blacklozenge$ and $\star$ symbols are for $\epsilon^*=6$.  All 
systems lie on the coexistence line except the $\star$ point is off 
coexistence at $\rho=0.83$, $\Phi=0.79\, m/\sigma^3$.  All elastic
constants are in units of $\sigma^5/m^2$.  The lines represent
fits given in Table~\ref{Ktable} and described in the text.}
\label{Ks}
\end{figure*}

The second derivatives of the free energy and the elastic constants can be
obtained from the linear fits in Fig.~\ref{Lmat}.  Fig.~\ref{d2fit} shows 
the second derivatives of $\psi$ as a function of $\rho$, and Fig.~\ref{Ks} 
shows the elastic constants plotted against $\rho$ and/or 
$\phi\equiv\Phi/\rho$.  Since values of $\psi$ and the $K$'s were 
only evaluated near the coexistence line, $\Phi_{co}(\rho)$, 
the dependence on density and concentration is intertwined.
The detailed procedure for finding the functional form of $\psi$ is discussed 
in Sect.~\ref{fits}.
We focus here on parameterizing the elastic constants because they are
important in understanding the interfacial tension results in the next
section.
Note that the variation of the $K$'s with $\rho$ and $\Phi$ was ignored
in the chemical potentials of Eqs. (\ref{murho}) and (\ref{muPhi}).  
Appendix~\ref{Kvar} explains why the variations discussed below can be ignored
in these equations. 

To separate the $\Phi$ and $\rho$ dependence of the elastic constants we
must consider points off the coexistence line. In principal this requires
simulations on a two-dimensional grid. However, we find that $K_\rho$ appears 
to depend only on $\rho$, while $K_{\rho\Phi}$ and $K_{\Phi}$ depend only on 
the scaled variable $\phi=\Phi/\rho$.  This is illustrated by including a 
point, $\rho=0.83 m/\sigma^3$, $\Phi=0.79m/\sigma^3$, in Fig.~\ref{Ks} 
that is away from the coexistence line.  This point collapses on the remaining 
data when $K_{\rho}$ is plotted against $\rho$ and $K_{\rho\Phi}$ and 
$K_{\Phi}$ are plotted against $\phi$.  However, it lies far from the 
other data when the elastic constants are plotted against the other variable, 
as illustrated in the plot of $K_\Phi$ vs. $\rho$.  Perhaps most surprising is 
that the results for different $\epsilon^*$ also collapse in Fig.~\ref{Ks} (a),
 (b) and (d).  Thus we can fit the data for the elastic constants with one
set of parameters.

Given the above results we take $K_\rho$ independent of $\Phi$ and fit it 
to a linear function of $\rho$ with the result given in Table~\ref{Ktable}.
By symmetry, $K_{\rho\Phi}$ must be an odd function of $\Phi$.  The dashed 
line in Fig.~\ref{Ks}(b) corresponds to a fit of the data assuming 
$K_{\rho\Phi} \sim \phi$ whereas the solid line is a fit assuming 
$K_{\rho\Phi} \sim \phi^3$.  The cubic function yields a better fit, giving
the result listed in Table~\ref{Ktable}. 
This suggests that the common practice of neglecting terms involving 
$K_{\rho\Phi}$ is justified in situations where $\phi$ is small,
 such as near the critical point, but is not justified in the more typical case
considered here.

Symmetry constrains $K_\Phi$ to be an even function of $\phi$.
The dotted line in Fig.~\ref{Ks}(d) comes from a fit of $K_\Phi$ to
a quadratic function of $\phi$.  This clearly deviates systematically
from the data which appears to be approaching a much flatter function
of $\phi$ as $\phi \rightarrow 0$. The solid and dashed lines,
which both furnish good fits over the range of data, correspond
respectively to the $\phi^8$ and $\phi^{10}$ fits given in 
Table~\ref{Ktable}.  As we shall see in the next section, the value of 
$K_\Phi$ has a significant impact on interface properties.  However, these 
properties are most strongly dependent on the value of $K_\Phi$ around 
$\phi=0$.  As such, the fact that $K_\Phi$ is nearly constant in this region 
is a highly desirable result.  However, in order to have confidence in an 
extrapolation of $K_\Phi$ from values of $\phi > 0.74$ down to 
$\phi=0$ requires that we examine the interfaces explicitly.  

\begin{table}
\begin{tabular}{c|c}
\hline
\hline
%               & fit  \\ 
%\hline
$K_\rho$       &  $\,(3.16 \pm 0.5\, \frac{\sigma^5}{m^2})\,+\,(-6.08 \pm 0.57\,  \frac{\sigma^{8}}{m^3})\,\rho\,$   \\
\hline
$K_{\rho\Phi}$ & $(-0.225\pm0.011\,\frac{\sigma^{5}}{m^2})\,\phi^3$  \\
\hline
$K_{\Phi}$     & $\,(0.286 \pm 0.015\, \frac{\sigma^5}{m^2})+(0.517 \pm 0.031\, \frac{\sigma^5}{m^2})\,\phi^8 \,$  \\
               &  $\,(0.319 \pm 0.013\, \frac{\sigma^5}{m^2})+(0.524 \pm 0.03\, \frac{\sigma^5}{m^2})\,\phi^{10} \,$ \\
\hline
\hline
\end{tabular} 
\caption{Elastic constants measured for $\epsilon^*=5$ and $6$ and $0.82\, m/\sigma^3 < \rho < 0.925 \,m/\sigma^3$.  Care should be taken in extrapolating the
elastic constants outside of the measured range.  The two alternatives given
for $K_\Phi$ fit equally well over the range of the linear response data 
($0.74<\phi <0.97$).
\label{Ktable}}
\end{table}

\subsection{Interface characterization}
\label{MDinterface}

\begin{figure}
\includegraphics[width=3.1in]{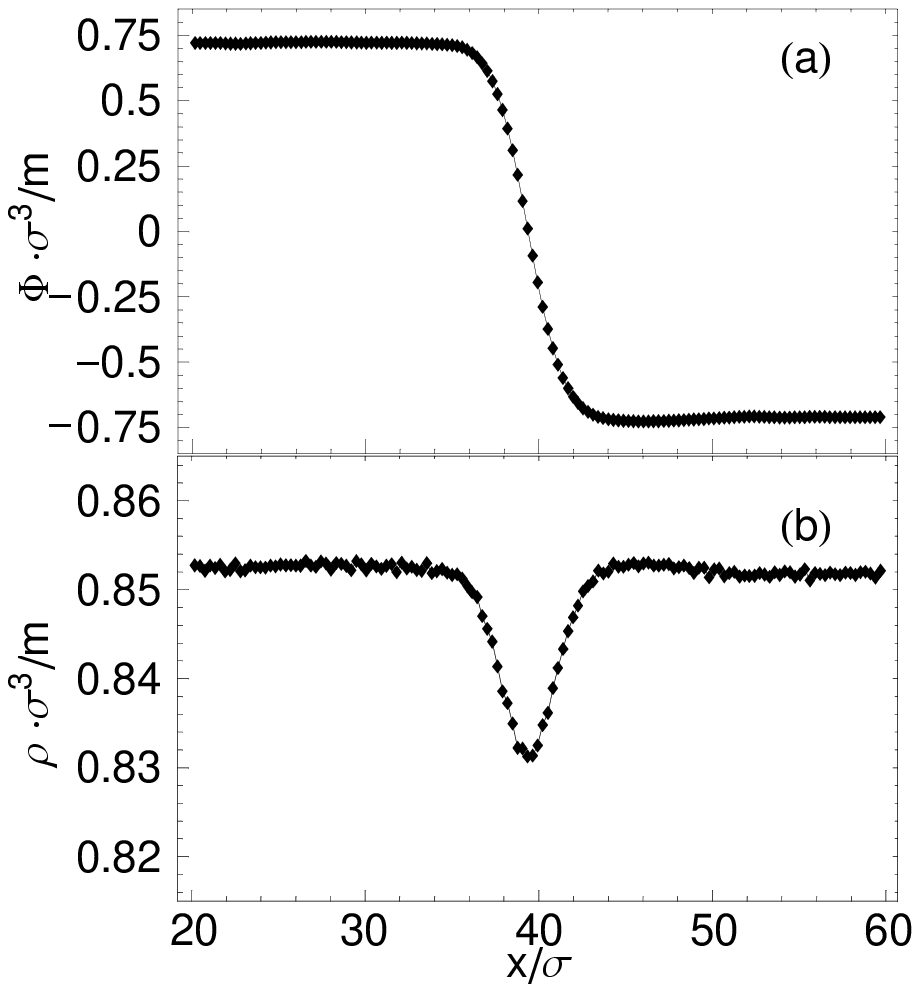}
\vskip 0.2true cm
\includegraphics[width=3.1in]{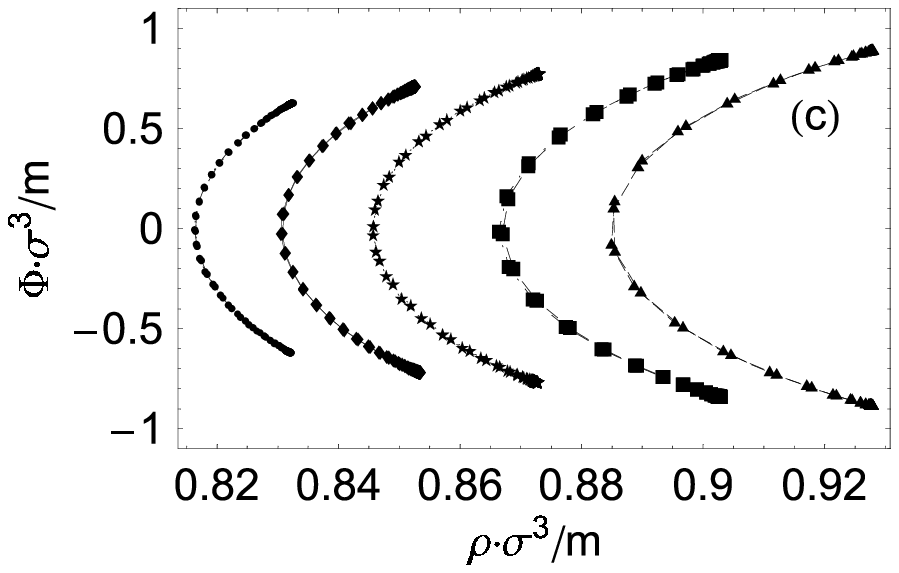}
\caption{(a) Order parameter $\Phi$ and (b) density $\rho$ along the
  $x$-direction of the simulation box, and averaged over the $y$ and
  $z$ directions.  The temperature is $T=0.8$ and $\epsilon^*=5$. 
  (c) $\Phi$ versus $\rho$ for different interfaces with $T=1.1$ 
  and average density of 0.83 ($\bullet$), 0.85 ($\blacklozenge$), 
0.87 ($\star$), 0.9 ($\blacksquare$), and 0.925 ($\blacktriangle$).}
\label{Phirho}
\end{figure}

We will examine the geometry shown in Figure~\ref{atompic}.  There are
periodic boundary conditions in all directions and two flat (on
average) interfaces normal to the $x$-axis.  We have examined different 
runs with average densities ranging from $\rho=0.82 m/\sigma^3$ to
$\rho=0.925 m/\sigma^3$, temperatures from $k_B T/\epsilon=0.8$ to 
$k_B T/\epsilon=2.0$, and $\epsilon^*$ from 
$5$ to $8$.
% , and with $\sigma_{ab}$ of $1.1 \,\sigma$ and $1.0 \,\sigma$.  
Fig.~\ref{Phirho} shows $\Phi$ and $\rho$ for a typical system at 
$k_B T/\epsilon=0.8$.   

Most models of binary fluids in the literature assume incompressibility
and therefore that $\rho$ is constant.  This is a reasonable assumption in 
the bulk but is violated at an interface, as illustrated in Fig.~\ref{Phirho}.
 The density dip in panel (b) is a 
ubiquitous feature of fluid-fluid interfaces noted recently in several 
publications \cite{TS95,DH99}, including some involving polymer 
fluids \cite{MRbar,G98}.  The dip occurs because the energetically unfavorable 
$1$-$2$ particle interactions are concentrated at the interface.  The
system can lower its free energy by decreasing the overall density at
the interface as this reduces the number of $1$-$2$ particle interactions. 
The size of the dip is determined by balancing 
this energy gain against the entropic cost from denying particles 
at the interface some volume. 

The surface tension associated with the interface is the most important 
characteristic for determining macroscopic behavior.  
The mechanical definition due to Kirkwood and Buff \cite{KB49} relates
the surface tension to the integral of the difference between the normal 
$P_\perp$ and parallel $P_\parallel$ components of the pressure tensor across 
the interface. 
For a flat interface normal to the $x$-axis, this can be written as
\begin{equation}
\gamma=\int \left[ P_{xx}(x)-(P_{yy}(x)+P_{zz}(x))/2 \right] dx.
\label{KB}
\end{equation}
In the quiescent state, $P_{xx}$ is constant throughout the system 
(no flow implies $\partial_\alpha P_{\alpha\beta}=0$ and all quantities 
are constant in both the $y$ and $z$ directions).  

Far from the interface the pressure is isotropic and one expects
$P_{xx}=P_{yy}=P_{zz}$.  However, for small systems this expectation
fails.  For example for $L_y=L_z=8.2 \sigma$ and 
$L_x=37.8 \sigma$ we found that $P_\perp-P_\parallel=0.004$ in a 
homogeneous system (i.e. without interfaces).  Although this sounds
small, when it is integrated over the whole system it yields 
$(P_\perp-P_\parallel)L_\perp=0.15$.  This integrated stress difference is
not related to the surface tension but still gives a significant
contribution to Eq.(\ref{KB}).  
The effect appears to be due to the fact that the density-density correlation 
function has not decayed to zero at a separation of $L_y=L_z$ but has
by $L_x$ \cite{DR02a}.  This appears to be a significant 
unrecognized error in molecular dynamics calculations of surface tensions.
For the larger systems for which we 
present data, we find that $(P_\perp-P_\parallel)\leq 0.0002$ and
$(P_\perp-P_\parallel)L_\perp\leq 0.015$.

\begin{figure}
\includegraphics[width=3.2in]{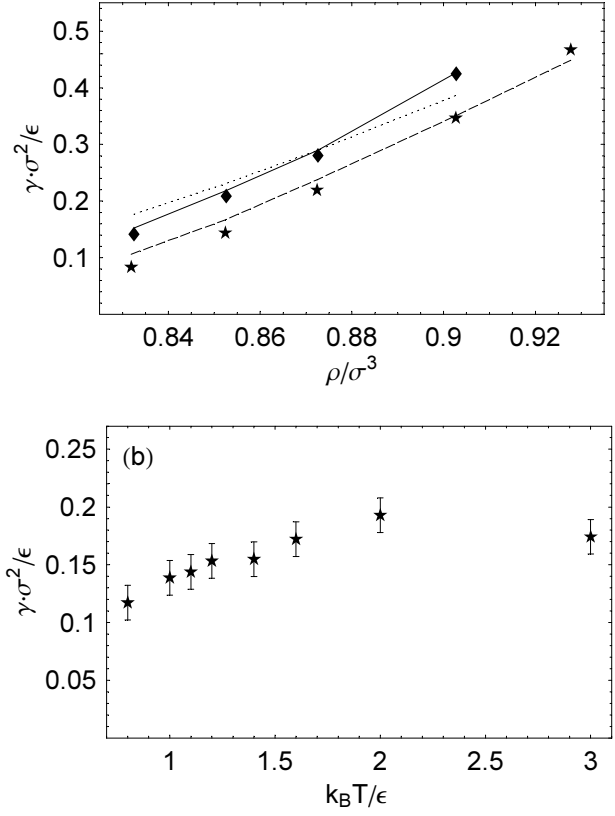}
\caption{The surface tension $\gamma$  for 
$\epsilon^*=5$ ($\star$) and $6$ ($\blacklozenge$)
as a function of (a) average density
at $T=1.1$ and (b) temperature at $\rho=0.85$.  Error bars 
represent systematic errors from $\langle P_{xx}-(P_{yy}+P_{zz})/2\rangle$ 
in the bulk regions.  The dashed and solid lines in (a) are the 
surface tension computed using the lattice Boltzmann method for $\epsilon^*=5$ 
and $6$, respectively, as discussed in Section \ref{LBresults}.  The dotted
line is the surface tension computed using the lattice Boltzmann method with
the Flory-Huggins free energy discussed in Appendix \ref{psivar} for 
$\epsilon^*=6$.}
\label{stension}
\end{figure}

Fig.~\ref{stension} shows measured surface tensions as a function
 of density, $\epsilon^*$, and temperature.  
Normally, one expects the surface tension to decrease as you
move towards the critical point.  This is certainly the case
as one decreases the density (Fig.~\ref{stension}(a)).  However,
the surface tension increases as a function of temperature
despite the fact that increasing the temperature moves the system
towards the critical point.  This effect has been noted before
by a number of workers \cite{IF97,DH99}.  The surface tension 
peaks around $k_B T/\epsilon=2$ and then starts 
decreasing \cite{DH99}.  As one increases $T$, there is a stronger mixing
at the interface that introduces more of the weaker $1-2$ interactions
and raises the potential energy.  This leads to an increase in $\gamma$
until at higher temperatures the entropy contribution brings it back
down again \cite{DH99}.

\begin{figure}
\includegraphics[width=3.2in]{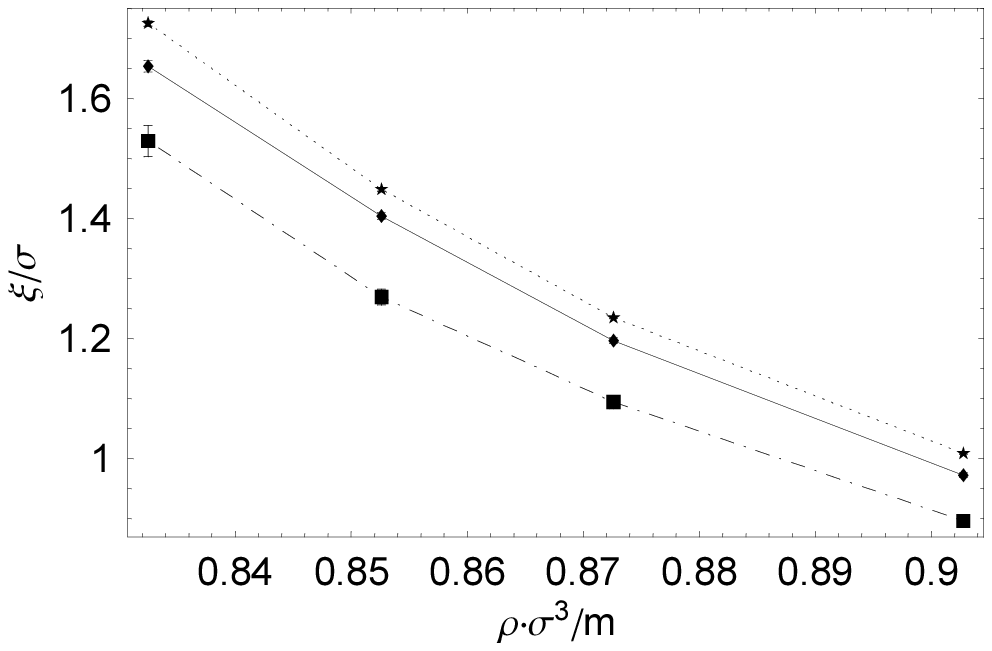}
\caption{The interface width for $\epsilon^*=6$ as a function of average 
density calculated from fits of the order parameter $\Phi$ profile to an 
error function ($\blacklozenge$) and the density profile to a constant minus a
Gaussian ($\star$).   The width obtained from the stress profile $\Gamma$
is also shown ($\blacksquare$).  Lines are only a guide to the eye.}
\label{intwidth}
\end{figure}

In addition to the surface tension, it is possible to define a characteristic
width to the interface.  The interface profile for $\Phi$ from the molecular 
dynamics simulations (Fig.~\ref{Phirho}(a)) can be fit to a $\tanh$ shape or 
equally well to an error function ${\rm erf}((x-x_0)/\sqrt{2 \xi^2})$.  
Similarly $\rho$ can be fit to a constant minus a Gaussian, 
$A \exp((x-x_0)^2/(2 \xi^2))$.  The resulting widths $\xi$ are shown in 
Fig.~\ref{intwidth}.  

\begin{figure}
\includegraphics[width=3.2in]{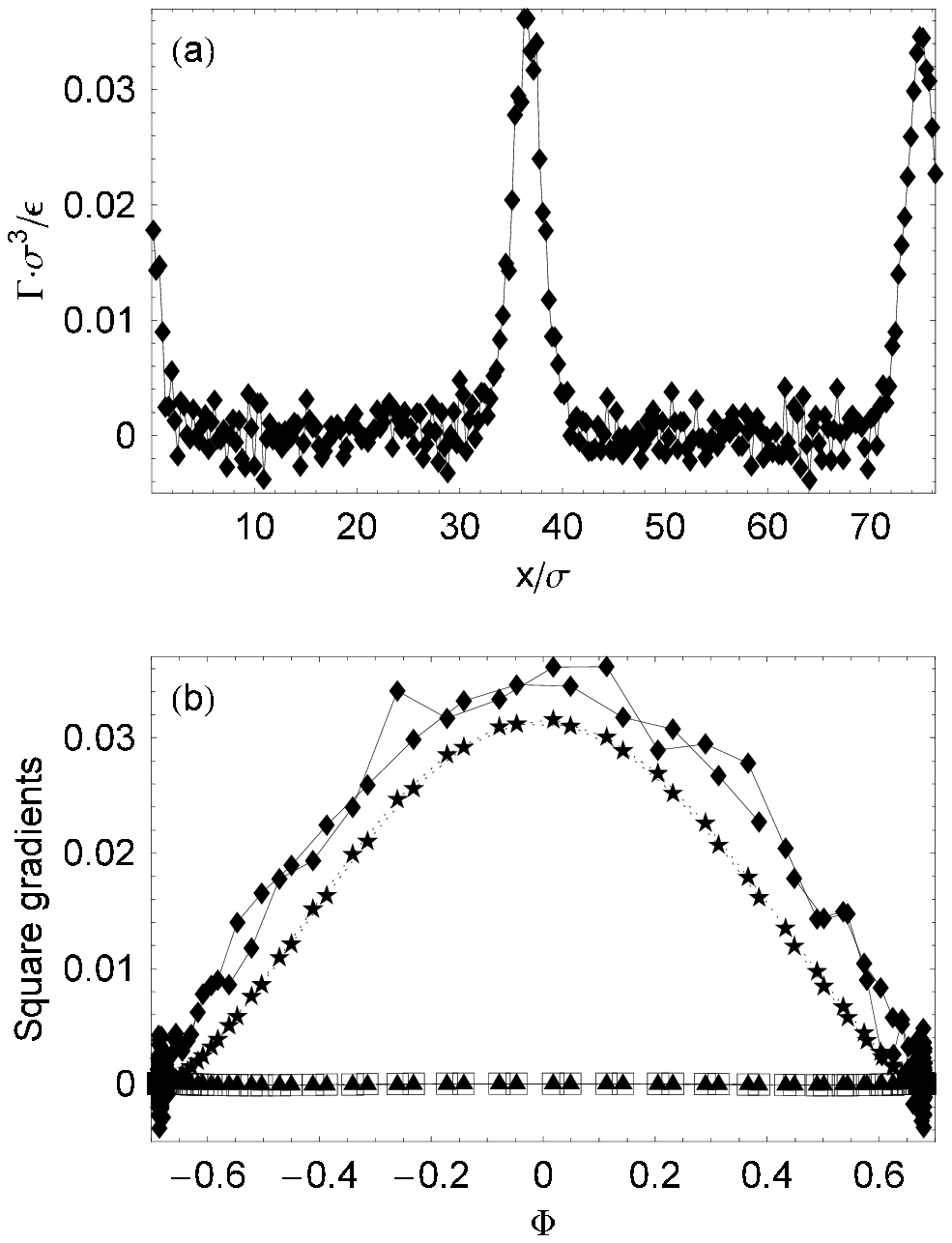}
\caption{(a) The interface stress difference 
$\Gamma(x)=P_{xx}(x)-(P_{yy}(x)+P_{zz}(x))/2$ for a system with average
density $\rho=0.83$ and $\epsilon^*=6$.
(b) The interface stress difference $\Gamma$ ($\blacklozenge$) and
square gradient contributions $K_\Phi (\partial_x \Phi)^2$ ($\star$), 
$K_\rho (\partial_x \rho)^2$ ($\Box$), and 
$K_{\rho\Phi}(\partial_x \rho)(\partial_x \Phi)$ ($\blacktriangle$).
The elastic constants were taken from the linear response data (the $\phi^{8}$
fit was used for $K_\Phi$).  Data is for a system with average density $0.83$
and $\epsilon^* =6$.  Note that due to the periodic boundary conditions there
are two interfaces in the system, giving two independent curves in (b).}
\label{Gamfig}
\end{figure}

The stress profile through the interface also provides useful information.  
The surface tension is related to the stress through the kernel of 
Eq.(\ref{KB}), 
\begin{equation}
\Gamma(x)\equiv P_{xx}(x)-(P_{yy}(x)+P_{zz}(x))/2, 
\label{GammaP}
\end{equation}
We measure the local components of the stress tensor directly in the molecular 
dynamics simulations.  A typical example is shown in Fig.~\ref{Gamfig}(a).  
The variation of $\Gamma$ is also fit to a Gaussian to obtain a width.  
As shown in Fig.~\ref{intwidth}, this width is smaller than those obtained 
from the interfacial profiles of the $\rho$ and $\Phi$.  We will discuss this
difference below.

The interfacial stress profile is also related
to gradients of the density and order parameter.  Substituting the expression
for the pressure tensor Eq.(\ref{PressureTensor}) into Eq.(\ref{GammaP}) 
gives, 
\begin{equation}
\Gamma(x)= k_B T \left[K_\rho (\partial_x \rho)^2+K_\Phi (\partial_x \Phi)^2 +2 K_{\rho\Phi}(\partial_x \rho) (\partial_x \Phi)\right].
\label{Gammasq}
\end{equation}
The numerical derivatives 
 are computed by first doing a local quadratic fit to the data 
($\rho$ or $\Phi$) over a range of five to seven points (data was collected 
in bins with width $0.29 \sigma$) and then using the derivative of the local 
fitting function.  
Figure \ref{Gamfig}(b) shows the contributions from the various terms in 
Eq.~\ref{Gammasq}.  As is evident from the figure, the only significant
contribution to the interfacial stress comes from the 
$K_\Phi (\partial_x \Phi)^2$ term.  As a result, since the interface profile
for $\Phi$ is well fit by the error function    
${\rm erf}((x-x_0)/\sqrt{2 \xi^2})$ we would expect the width obtained
from a fit of the stress difference $\Gamma(x)$ to a Gaussian to give
a width $1/\sqrt{2}\approx 0.71$ of that from the fit to $\Phi$.  
The width we actually measure from $\Gamma(x)$ is about $0.9$ of that 
obtained from $\Phi$ (see Fig.~\ref{intwidth}).  That is, the stress
profile is wider than expected, a fact that is also obvious from
Fig.~\ref{Gamfig}(b).  This may be due to a failure of the square
gradient theory or may be a result of interface fluctuations not yet taken 
into account, namely capillary waves.

The widths measured in molecular dynamics simulations are not the
intrinsic values but widths broadened by thermal fluctuations.  
Capillary waves result in the measured time averaged width $\xi$ being larger 
than the intrinsic width $\xi_0$ by \cite{LGL98}
\begin{equation}
\xi^2=\xi_0^2+\frac{k_B T}{2 \pi \gamma}\ln\left(\frac{L}{\Delta_0}\right),
\label{capwidth}
\end{equation}
where $\Delta_0$ is a short scale cutoff.  $\Delta_0$ is expected to be
proportional to the intrinsic interface width (fluctuations on length scales 
shorter than the interface width are not capillary waves).  We take 
$\Delta_0\equiv c \xi_0$ where $\xi_0$ is the intrinsic width of
the $\Phi$ profile and $c$ is a numerical constant. The bare width from
the $\Phi$ profile is used for $\xi_0$ since the width from $\Gamma$ is a
derived quantity in the square gradient theory.  As all widths in the
problem are proportional to the intrinsic width of the $\Phi$ profile,
changing the width used in defining $\Delta_0$ will change $c$ but not 
$\Delta_0$.

\begin{figure}
\includegraphics[width=3.2in]{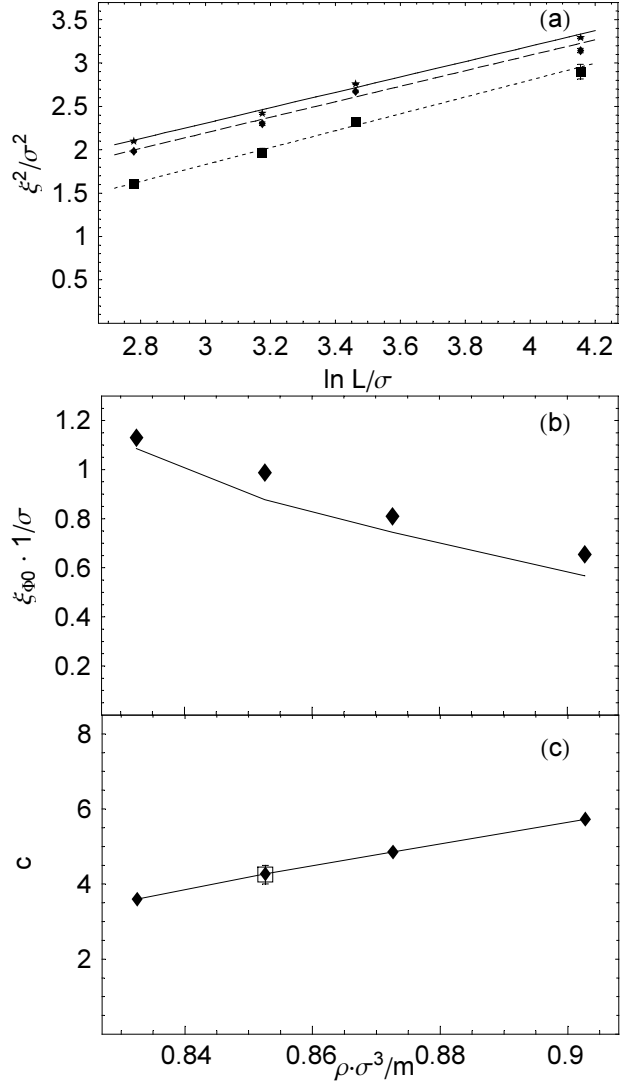}
\caption{
(a) Dependence of the interfacial widths computed from the $\Phi$ profile
($\blacklozenge$), $\rho$ profile ($\star$), and $\Gamma$ ($\blacksquare$) as a
function of the system size parallel to the interface.  The density is $0.85$
and $\epsilon^*=6$ for all systems.  Lines show linear fits to 
Eq.~\ref{capwidth}.
(b) The intrinsic interface width as a function of average density 
for the order parameter $\Phi$ profile from molecular
dynamics ($\blacklozenge$) and from lattice Boltzmann (line).  
(c) Parameter $c$ relating the short scale cutoff $\Delta_0$ to $\xi_{\Phi_0}$ 
calculated using Eq.(\ref{intwidtheq}) ($\blacklozenge$) and 
Eq. (\ref{adelta}) ($\square$).
Data for $\epsilon^*=6$ is shown but similar results are obtained for
 $\epsilon^*=5$.}
\label{intrinsicwidth}
\end{figure}

Following Ref.\cite{LGL98} we verify Eq.(\ref{capwidth}) by 
plotting the measured width as a function of system size.  This
is shown in Fig.~\ref{intrinsicwidth}(a).  Fitting the data to the
expression $\xi^2=a_\delta+b_\delta \ln L$, we find that 
$b_\delta=k_B T/(2 \pi \gamma)$ to
well within measurement errors, as expected from Eq.(\ref{capwidth}).
The other parameter from the fit 
\begin{equation}
a_\delta=\xi_0^2-b_\delta \ln c \xi_0
\label{adelta}
\end{equation}
could be used to obtain $\xi_0$ {\it if} we knew $c$.  Unfortunately
that information is {\it not} available from this fit alone.  In
Ref.\cite{LGL98}, $c \equiv 13$ was picked arbitrarily as the smallest
number where Eq.(\ref{adelta}) had a solution for all the systems
considered, including monomers and polymers with up to 30 monomers.  There
is, however, no reason to believe that $c$ should be a universal number
in this sense.

If we assume that square gradient theory describes the interfaces, 
just as it works for linear response, then the value of $c$ can be obtained
from the difference in the widths of the $\Gamma$ and $\Phi$
 profiles.  In the square gradient theory the intrinsic widths should obey
the relation $\xi_{\Gamma_0}^2=\xi_{\Phi_0}^2/2$,
where $\xi_{\Gamma_0}$ is the intrinsic width of $\Gamma$ and $\xi_{\Phi_0}$
the intrinsic width of $\Phi$.  Based on Eq.(\ref{capwidth}), the measured 
widths, $\xi_\Phi$ and $\xi_\Gamma$, of the $\Phi$ and $\Gamma$ profiles
should be
\begin{eqnarray}
\xi_\Phi^2 &=& \xi_{\Phi_0}^2+
\frac{k_B T}{2 \pi \gamma}\ln\left(\frac{L}{c \xi_{\Phi_0}}\right),\nonumber\\
\xi_\Gamma^2 &=& \frac{1}{2}\xi_{\Phi_0}^2+
\frac{k_B T}{2 \pi \gamma}\ln\left(\frac{L}{c \xi_{\Phi_0}}\right).
\label{intwidtheq}
\end{eqnarray}
Solving these equations simultaneously for $c$ and $\xi_{\Phi_0}$ gives 
the result shown in Fig.\ref{intrinsicwidth}(b).  Using Eq.(\ref{adelta}), we 
can also compute $c$ from the finite size scaling if we use the intrinsic 
widths obtained from Eq.(\ref{intwidtheq}).  The
agreement seen between these two methods in Fig.~\ref{intrinsicwidth}(c) 
gives weight to the assumption that square gradient theory 
describes the interfaces accurately. 

We can also assess the consistency of the values of $K_\Phi$ obtained
from the linear response with estimates from interface data.  To do this,
we assume capillary waves broaden the interface in a Gaussian manner (i.e. 
the measured shape of the interface is a convolution of the intrinsic shape 
and a Gaussian distribution of width $\sqrt{\xi^2-\xi_0^2}$).
If we further assume that $K_\Phi$ is constant and that the intrinsic line 
shape for $\Phi$ is reasonably approximated by an error function 
(which is the case for the broadened shape directly measured), then
\begin{equation}
\left[\frac{\Gamma(x=0)}{\partial_x \Phi (x=0)}\right]_{MD}=k_B T K_\Phi \frac{\xi_\Phi^2}{\sqrt{2}\xi_\Gamma \xi_{\Phi_0}}.
\end{equation}
The quantity on the left is the directly measured (capillary broadened)
stress difference divided by the gradient of $\Phi$ at the point where $\Phi$ 
crosses zero (center of the interface).  The ratio of the length scales,
$\xi_\Phi^2/(\sqrt{2}\xi_\Gamma \xi_{\Phi_0})$, is about $1.21\pm 0.05$ for the
 interfaces measured.  This gives a value of $K_\Phi$
of $0.29\pm 0.03 \sigma^5/m^2$.  As $\Gamma$ (and $\partial_x \Phi$) is 
peaked at $\Phi=0$ and goes to zero as $\Phi \rightarrow \Phi_{co}$ this 
measure of $K_\Phi$ is dominated by the value of $K_\Phi$ at $\Phi=0$.  In the 
previous section we determined that $K_\Phi$ is essentially constant near 
$\Phi=0$ and that in this region it should have a value around $0.286$ to 
$0.319 \, \sigma^5/m^2$.
Thus we can be reasonably happy that the elastic constants determined
from linear response are consistent with those obtained from the interfaces.

\section{Free energy parameterization}
\label{fits}

In order to use the information from the molecular dynamics simulations
in a mesoscopic model,  we also need to parameterize the bulk free energy 
density $\psi$.  As we have direct information about $\psi$ and its 
various derivatives from the interfacial stress and linear response data, 
this would not at first appear to be a difficult thing to do.  
However, $\psi$ is a function of both $\rho$ and $\Phi$.  
The addition of the order parameter means that expressions for obtaining 
the free energy from the pressure found in standard references \cite{AT87} 
are {\it not} applicable.  Fig.~\ref{Phirho}(c) shows that as one goes through 
an interface, $\Phi$ can be reasonably approximated by a quadratic function of 
the local density $\rho$.  As a result, distinguishing $\psi$'s dependence on 
$\Phi$ from that on $\rho$ using interfacial data alone is extremely
difficult.  Also, capillary waves change the shape of the interfacial
profiles so that getting information about $\psi$ directly from the interface
shape can be misleading.  We therefore use the linear response data to 
do the fits and then compare the resulting surface properties as a test of 
the parameterization in Section~\ref{LBresults}.

Before deciding on a particular functional form, it is worthwhile to 
examine what we would like to fit, and to prioritize what is most/least 
important.  Our fit priorities are:
\begin{enumerate}
\item[1.]
Phase diagram/coexistence line ($\Phi_{co}(\rho)$).
\item[2.]
Linear response of equilibrium phase (compressibility, etc).
\item[3.]
Surface tension.
\item[4.]
Interface widths.
\end{enumerate}
The rationale for this choice is fairly simple.  Any model incapable of 
satisfying the first item has little, if any, useful properties.  A model
capable of reproducing the first two items will at least have a reasonable
response to bulk forces.  If a model fits the first three items then it
should reproduce most macroscopic interfacial behavior.  If in addition the 
fourth item can be fit, then we can reasonably expect it to describe a number 
of microscopic processes with macroscopic consequences, such as droplet 
coalescence or pinch-off.

There are a number of free energy functionals commonly used
to study fluid mixtures.  We found Flory-Huggins theory has too few
parameters to satisfy our first two requirements, while a Landau-Ginzburg
expansion requires too many, resulting in ambiguous fits.  In this section
we discuss an alternative parameterization and relegate discussions of 
Flory-Huggins theory to Appendix~\ref{psivar}.

To minimize degeneracies in the fitting procedure, we will try to 
parameterize the free energy in terms of quantities that are directly measured
in the molecular dynamics simulations.   
The partition function we use is a sum of Gaussians centered at 
$\pm \Phi_{co}$ plus a third Gaussian centered at $\Phi=0$ whose amplitude
is chosen so as to partially cancel the overlap of the other two Gaussians.  
This results in the free energy density:
\begin{eqnarray}
k_B T\psi&=& A_0-\rho k_B T \ln\left[ 
\exp(-A_2(\Phi-\Phi_{co})^2/(2\rho k_B T)) \right.\nonumber\\
&& + \exp(-A_2(\Phi+\Phi_{co})^2/(2\rho k_B T)) \nonumber\\
&& \left.- \exp(-A_2(\Phi^2+\Phi_{co}^2)/(\rho k_B T))\right]
\label{psifit}
\end{eqnarray}
where $A_0$, $A_2$, and $\Phi_{co}$ are functions of $\rho$ to be
determined. As we will show, Eq.(\ref{psifit}) reproduces the linear 
response data to within statistical errors essentially by construction.
For a single phase system on the coexistence line, but far from the critical  
point so $\Phi_{co}>> 0$, Eq.(\ref{psifit}) simplifies to a more familiar form
$$
k_B T \psi=A_0+\frac{1}{2}A_2(\Phi-\Phi_{co})^2.
$$

The parameters of the free energy functional can be
determined directly from our molecular dynamics measurements of the
equilibrium state.
On the coexistence line ($\Phi=\pm \Phi_{co}$) one can easily show that
\begin{eqnarray}
\frac{\partial^2 \psi}{\partial \rho^2}&=&\frac{1}{k_B T} \frac{\partial^2 A_0}{\partial \rho^2}+\frac{\partial^2 \psi}{\partial \Phi^2}\left(\frac{d \Phi_{co}}{d \rho}\right)^2,\label{D2psirho2}\\
\frac{\partial^2 \psi}{\partial \rho \partial \Phi}&=& -\frac{\partial^2 \psi}{\partial \Phi^2}\left(\frac{d \Phi_{co}}{d \rho}\right),\label{D2psirhoPhi}\\
\frac{\partial^2 \psi}{\partial \Phi^2}&=& \frac{A_2}{k_B T} \left(1-\exp\left(-\frac{2 A_2 \Phi_{co}^2}{\rho k_B T}\right)\right).\label{D2psiPhi2}
\end{eqnarray}
For states where $\Phi_{co}$ is significantly greater than zero the last
expression reduces to 
$\partial^2 \psi /\partial \Phi^2= A_2$.
The equilibrium hydrostatic pressure is
\begin{eqnarray}
p_0=\rho \frac{\partial A_0}{\partial \rho}-A_0.
\label{p0ofA0}
\end{eqnarray}
By rearranging these equations it is also straightforward to show that
along the coexistence line
\begin{eqnarray}
\frac{d p_0}{d \rho} &=& \rho k_B T \left[\frac{\partial^2 \psi}{\partial \rho^2}- \frac{\left(\frac{\partial^2 \psi}{\partial \rho \partial \Phi}\right)^2}{\frac{\partial^2 \psi}{\partial \Phi^2}} \right]=\rho \frac{\partial^2 A_0}{\partial \rho^2},\\
\frac{d \Phi_{co}}{d \rho} &=& -\frac{\frac{\partial^2 \psi}{\partial \rho \partial \Phi}}{\frac{\partial^2 \psi}{\partial \Phi^2}}.
\label{dsofA0}
\end{eqnarray}

\begin{figure}
\includegraphics[width=3.2in]{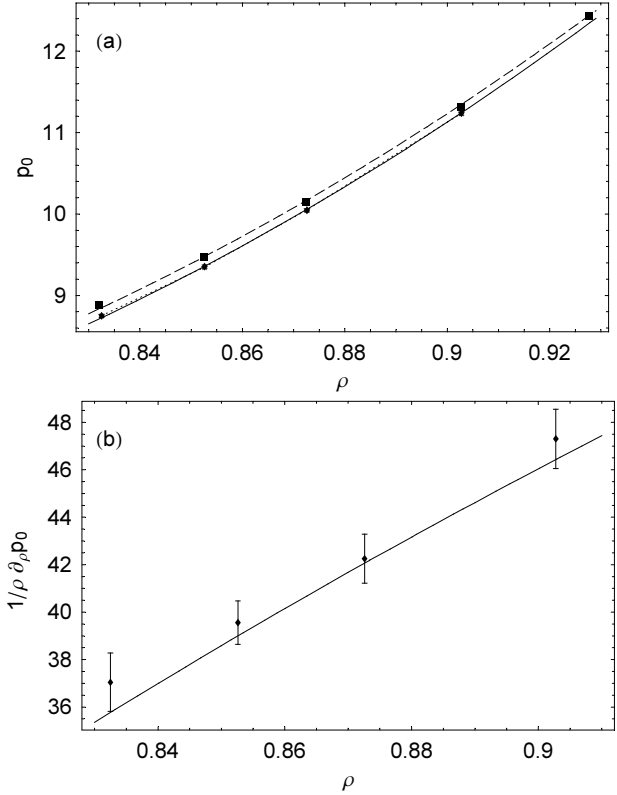}
\caption{Fits to (a) $p_0$ and (b) $dp_0/d\rho$ which give the parameterization
of $A_0$ in Eq.(\ref{A0form})).  In (a), molecular dynamics data is shown 
for $\epsilon^*=5$ ($\blacksquare$) and $\epsilon^*=6$ ($\blacklozenge$) and
the solid and dashed lines correspond to the fits described in the main text.
The dotted line in (a) (which overlaps with the solid line except at small 
$\rho$) corresponds to the fits used to determine $\psi_0$ for 
the Flory-Huggins model described in Appendix~\ref{psivar} for $\epsilon^*=6$.
In (b) only the $\epsilon^*=6$ data is shown.}
\label{p0fit}
\end{figure}

To use Eq.(\ref{psifit}) in the lattice Boltzmann algorithm, we need to choose 
a functional form for the dependence of the parameters
on $\rho$.  As we have information about both values and derivatives (of
$\Phi_{co}$ for instance), one could use a spline fit to match the measured
values exactly.  Alternatively, we can pick a functional form for the 
parameters and fit them over a range of $\rho$.  For $A_0$, a convenient
form is
\begin{equation}
A_0=a_{0}+a_{1} \rho \ln \rho + a_{2} \rho^2.
\label{A0form}
\end{equation}
Then from Eq.(\ref{p0ofA0}),
\begin{equation}
p_0= -a_{0}+a_{1} \rho + a_{2} \rho^2
\end{equation}
and from Eq.(\ref{dsofA0})
\begin{equation}
\frac{1}{\rho}\frac{dp_0}{d\rho}|_{co}=\frac{\partial^2 \psi}{\partial \rho^2}-\frac{\left(\frac{\partial^2 \psi}{\partial \rho \partial \Phi}\right)^2}{ \left(\frac{\partial^2 \psi}{\partial \Phi^2}\right)}=a_{1} \frac{1}{\rho}+2 a_{2}.
\end{equation}
A simultaneous fit of $p_0$ and $dp_0/d\rho$, weighted by the statistical 
errors from the measurements of $p_0$ and the second derivatives used to obtain
 $dp_0/d\rho$, is shown in Fig.\ref{p0fit}.  The values of the fit parameters
are given in Table~\ref{psitable}.

Similarly, it is straightforward to do a fit of $\Phi_{co}$ to a quadratic
function taking into account the information about $d\Phi_{co}/d\rho$ 
(Eq.(\ref{dsofA0})):
\begin{equation}
\Phi_{co}=b_0+b_{1} (\rho-\rho_0) + b_{2} (\rho-\rho_0)^2,
\label{phicoform}
\end{equation}
where $\rho_0=0.85\, m/\sigma^3$ is a reference density chosen so as to make
the statistical errors in the fit parameters given in Table~\ref{psitable}
independent.  This fit is shown in Fig.~\ref{rhophi}.
The parameterization of the remaining quantity, $A_2$, is done in two steps.  
First, Eq. (\ref{D2psiPhi2}) is numerically inverted to obtain $A_2$ at each 
point using the directly measured $\Phi_{co}$ and 
${\partial^2 \psi}/{\partial \Phi^2}$.  The resulting values are then
fit to the function:
\begin{equation}
A_2=c_0 \exp \left( c_{1} (\rho-\rho_0)\right).
\label{A2form}
\end{equation}
The fitted parameters are given in Table~\ref{psitable} and the resulting fit
to ${\partial^2 \psi}/{\partial \Phi^2}$ is shown in Fig.~\ref{d2fit}(c).  
An exponential was chosen to fit $A_2$ rather than a quadratic or other 
polynomial for two reasons.  First, it gave a better fit, and second, the 
quadratic did not remain monotonic over the full range of interest leading to 
unphysical effects when extrapolating outside the range of densities where 
linear response was measured.

One could, in principle, add additional terms to the bulk free energy density
that would have {\it no} impact on the coexistence curve or the linear 
response data fitted so far.   Such a term would be zero and have zero first
and second derivatives on the coexistence line.  If, in addition, it was peaked 
at $\phi=0$ it would affect the surface tension and interface width.  
As discussed in the next section, we do not need such terms here but they may 
be useful in other contexts.

\begin{table}
\begin{tabular}{c|c|c}
\hline
\hline  
$\epsilon^*$&                  5                        &            6     \\
\hline
$a_{0}$     & $-43.92 \pm 5.72\, \frac{m}{\sigma \tau^2}$ & $-43.78 \pm 5.8\, \frac{m}{\sigma \tau^2}$\\
$a_{1}$     & $-113.8\pm 13.0\,\frac{\sigma^2}{\tau^2}$  & $-114.0\pm 13.4\,\frac{\sigma^2}{\tau^2}$  \\
$a_{2}$     & $86.04\pm 7.42\,\frac{\sigma^5}{m \tau^2}$ & $86.37\pm 7.7\,\frac{\sigma^5}{m \tau^2}$  \\
\hline
$b_{0}$     & $0.695 \pm 0.004\, \frac{m}{\sigma^3}$ & $0.739 \pm 0.003\, \frac{m}{\sigma^3}$ \\
$b_{1}$     & $3.37 \pm 0.15\, $                     & $2.69 \pm 0.12\, $     \\
$b_{2}$     & $-12.0 \pm 1.6\, \frac{\sigma^3}{m}$   & $-9.5 \pm 1.6\, \frac{\sigma^3}{m}$    \\
\hline
$c_{0}$     & $2.011 \pm 0.060\,\frac{\sigma^5}{m \tau^2}$   & $2.827 \pm 0.007\,\frac{\sigma^5}{m \tau^2}$\\
$c_{1}$     & $20.40 \pm 0.67\,\frac{\sigma^3}{m}$ & $21.70 \pm 0.09\,\frac{\sigma^3}{m}$\\
\hline
\hline
\end{tabular} 
\caption{Table of parameters for the free energy $\psi$ in Eq. (\ref{psifit}).
  Parameters are defined in Eqs. (\ref{A0form}), 
(\ref{phicoform}), and (\ref{A2form}).  As for the elastic constants given in 
Table~\ref{Ktable}, the data were measured for 
$0.82\, m/\sigma^3 < \rho < 0.925 \,m/\sigma^3$ and care should be taken in 
extrapolating outside of the measured range. 
\label{psitable}}
\end{table}

\section{Tests of Parameterization}
\label{LBresults}

To test the above parameterization's description of interface properties we 
make use of lattice Boltzmann simulations.  
Lattice Boltzmann simulations were run for the same system size and 
compositions as the molecular dynamics simulations.  We first verified that 
the lattice Boltzmann program correctly reproduced the bulk data.  
Surface tensions were then computed from the integral of the stress difference
through the interface, Eq.(\ref{GammaP}), just as we did for the molecular
dynamics simulations.  Figure~\ref{stension} shows the surface 
tension computed from the lattice Boltzmann program using the elastic
constants and $\psi$ parameters from Table~\ref{psitable}.  Although the
parameterization only used the linear response data, the surface tensions are
 very close to the molecular dynamics results, although they tend to be 
slightly higher.  As discussed in Section~\ref{MDinterface},  the interfacial
properties are dominated by the peak in the free energy and value
of the elastic constants near $\phi=0$.  All the fits to the data 
were for values of $\phi > 0.73$, and it is remarkable that 
such an extrapolation works so well.

Due to capillary waves we cannot directly compare the full interfacial 
profiles of the density and order parameter to the molecular dynamics.
However, we can verify that the total mass deficit at the interface
is correct.  This can be calculated from the integral under the dip in plots 
like Fig.~\ref{Phirho}(b) or from the increase in the density far from the
interface that compensates for the dip.  For example, consider a molecular 
dynamics simulation where the total average system density is 
$\rho=0.85 m/\sigma^3$, $k_B T/\epsilon=1.1$, and with length 
$L_x=74.36 \sigma$.  We find that the density far from the interface 
$\rho=0.8526 \pm 0.0001 m/\sigma^3$ is larger than the system average
to make up for the deficit at the interface.  A lattice Boltzmann simulation
with the same system size and total average density shows exactly the
same mass increase in the bulk. 

Figure \ref{intrinsicwidth}(a) shows that there is also reasonable
agreement between the intrinsic widths computed via the lattice Boltzmann
and the molecular dynamics.  Although in this case the intrinsic widths
measured from the lattice Boltzmann tend to be a bit smaller.

To test our model parameterization in a different geometry we examined a 
cylindrical drop.  In three dimensions, a cylinder of fluid is unstable to 
spherical droplet formation.  We avoided this instability by making the
radius $R$ of the droplet bigger than the system size along the axis of
the cylinder ($y$-axis).  Specifically, $R=24.5 \sigma$, $L_y=16 \sigma$, 
and  $L_x$ and $L_z$ were about $100 \sigma$.  For a density
of $\rho=0.85 m/\sigma^3$ this required $147456$ molecules and $10^7$ time
steps were needed to get good statistics, which is 
reasonably large for a molecular dynamics simulation.  The only significant 
difference between the molecular dynamics simulation and
the lattice Boltzmann method is the presence of thermal noise.  In the 
molecular dynamics simulation this causes the drop to undergo Brownian
motion. In order to do a meaningful comparison we limited this effect by 
periodically shifting the system so that the center of mass of the
drop remained at $x=z=0$.  

\begin{figure}
\includegraphics[width=3.2in]{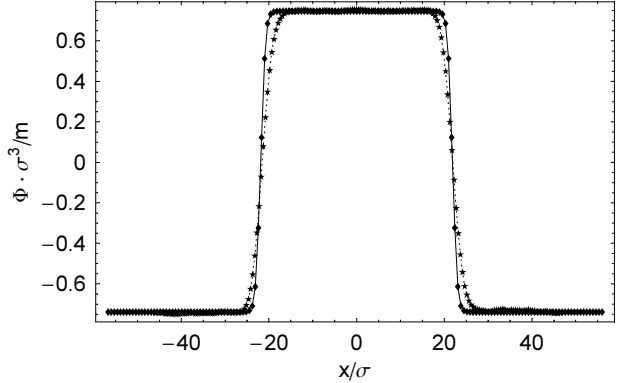}
\caption{Order parameter profile on a cut going through the center of
the drop.  Note that the molecular dynamics profile has been smeared
somewhat due to capillary waves and Brownian motion of the drop that 
are not present in the lattice Boltzmann simulation.}
\label{dropPhi}
\end{figure}

\begin{figure}
\includegraphics[width=3.2in]{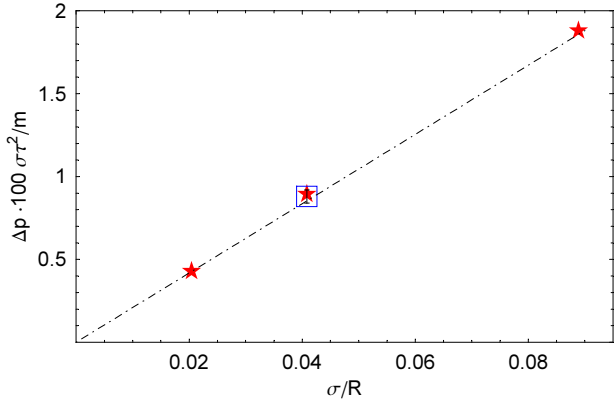}
\caption{Laplace pressure between the inside and
outside of a cylindrical fluid ``drop'', versus the curvature $1/R$, where
$R$ is the drop radius.  Results from lattice Boltzmann simulations ($\star$) 
and molecular dynamics simulations ($\square$) are consistent with each other
and with the straight line prediction from continuum theory Eq.\ref{laplace}.}
\label{laplaceP}
\end{figure}

Figure~\ref{dropPhi} compares profiles of the order parameter through the
center of the drop from molecular dynamics and lattice Boltzmann simulations.
The molecular dynamics profile is broadened slightly by capillary waves and
Brownian motion.  Neither effect is present in the lattice Boltzmann 
simulation.  The two methods both show that $\Phi$ is lower by 
$0.01 m/\sigma^3$ than the equilibrium bulk value in the center of the
drop.  The reason for the drop in $\Phi$ is that the interfacial curvature
produces a pressure difference $\Delta p_0$ between the inside and outside 
called the Laplace pressure.  This also increases the density in the drop.  
For $R=24.5 \sigma$ the density difference from lattice Boltzmann simulations
is $0.00867 m/\sigma^3$, which agrees with the value of
 $0.009 \pm 0.002 m/\sigma^3$ from molecular dynamics.  For a macroscopic
cylinder, the pressure difference between the inside and outside of the drop 
should scale with the radius as \cite{LL87}
\begin{equation}
\Delta p_0= \gamma /R,
\label{laplace}
\end{equation}
 where $\gamma$ is the surface tension measured in the molecular dynamics 
simulation of a flat interface.  As can be seen in Figure~\ref{laplaceP}, both
the molecular dynamics simulation and the lattice Boltzmann results follow this
relationship very well.

\section{Model Comparisons}

It is worthwhile to compare our findings to commonly used assumptions
in mesoscale modeling of binary fluids.  We first consider the assumption of 
incompressibility that is often used for both simple and binary fluids. 
For bulk fluids, including those examined here, this is a very reasonable
assumption as $\partial^2 \psi/\partial \rho^2$ is very large compared
to other second derivatives of the free energy (cf. Section~\ref{linresponse}).
However, due to the fact that $K_\rho$ is negative, the fluid can be very
soft on the short length scales characteristic of interface widths.
This invalidates the assumption of incompressibility in the interfacial
region. 

As noted in Section~\ref{mesomod}, the density drops in the interfacial region
to reduce unfavorable $1-2$ 
molecular interactions and hence the free energy.
The quantitative impact of density changes can be seen by considering
the change in the free energy barrier (Eq. \ref{psifit}) in
the interfacial region.
Due to the change in density in the interface, we find that $A_2$ drops
down to around 50 percent of its value in the bulk.
This reduction in $A_2$ 
spreads the Gaussian functions in Eq.(\ref{psifit}) thereby reducing
the barrier between the $\pm \Phi_{co}$ states.
%MR Give a factor for the reduction?
There is a corresponding reduction in the total interfacial tension
which is given by the integral of $\Gamma(x)$
through the interface (Eqs. \ref{KB} and \ref{GammaP}).
Using the Euler-Lagrange equations (Eqs. \ref{murho} and \ref{muPhi})
one can show that $\Gamma(x)$ is directly
related to the free energy barrier: $\Gamma(x)=k_B T \psi -A_0$.
To see the impact of density changes in the interface we can compare the
integral of $\Gamma(x)$ using the $\Phi$ profile from the lattice Boltzmann
simulation combined with either a $\rho$-dependent value of $A_2$
or a constant bulk value of $A_2$.
We find that including density variations reduces the surface 
tension by a factor of $2-4$ relative to incompressible models.   
Thus the density drop at the interface 
is not a negligible effect for any quantitative study 
involving interfaces.

\begin{figure}
\includegraphics[width=3.1in]{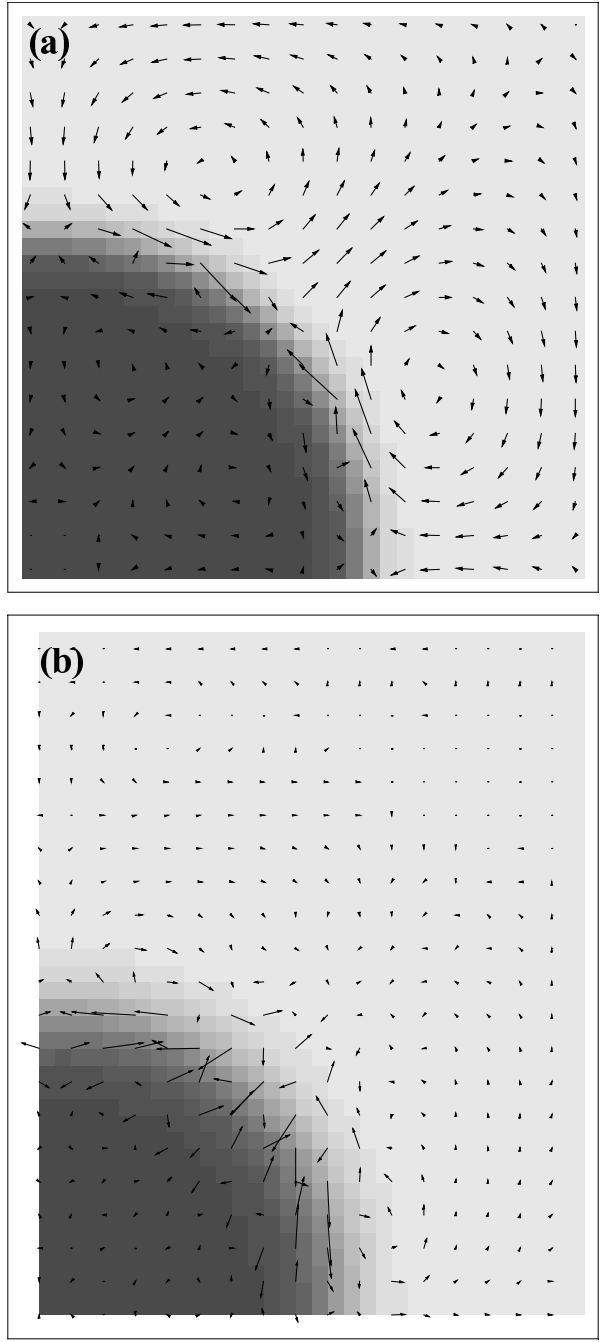}
\caption{$\Phi$ field (shading) and velocity field (vectors) for one quadrant
of a stationary cylindrical fluid ``drop''.  In (a), a standard model from the 
literature \cite{SY96} was used and in (b), the model we have matched to the 
molecular dynamics data. 
 In units where $dx=dt=1.0$, the largest velocity vector is 
$1.2\times 10^{-4}$ in (a) and $1.5 \times 10^{-6}$ in (b).  Note that the 
velocities in (b) would not be visible if we had used the same scale as that 
used in (a).  The radius of the drop in (b) is slightly smaller but this 
should only increase spurious velocities compared to (a).}
\label{spurvel}
\end{figure}

As mentioned before, the Flory-Huggins theory has too few parameters to 
obtain a precise fit to both the coexistence curve and the linear response.
As discussed in Appendix \ref{psivar}, we can obtain the Flory-Huggins
$\chi$ parameter from a fit to the coexisence curve shown in Fig.~\ref{rhophi}.
The resulting linear response seen in Fig.~\ref{d2fit}, while not in 
quantitative agreement with MD results,
should be adequate for qualitative work.  The
Flory-Huggins prediction for the surface tensions seen in Fig.~\ref{stension}
is actually remarkably good.  It is important to note however, that we have
allowed $\chi$ to vary with the local density.  As a result, $\chi$ drops
significantly in the interfacial region due to the interfacial density drop.  
This reduces the surface tension by a large factor compared to incompressible 
models, just as the reduction in $A_2$ discussed above did.  As most 
implementations of Flory-Huggins models for simulations assume 
incompressibility, such models will have surface tensions that
are a factor of $2-4$ too large.

It is commonly found in lattice Boltzmann simulations that a stationary 
droplet in quiescent conditions will develop a flow field around it similar
to that shown in Fig.~\ref{spurvel}(a) \cite{SY96,AW02}.  It has recently been
pointed out that these spurious velocities arise due to discretization errors
\cite{AW02} which drive the spurious currents.  For our model, fitted to
the molecular dynamics data, we can estimate the truncation error from
the discretization to be $\sim 10^{-6}$ and the velocities seen at the 
interface in Fig.~\ref{spurvel}(b) are indeed of this magnitude (in lattice
units).  The discretization error for the standard model from the literature
used to produce Fig.~\ref{spurvel}(a) should also be $\sim 10^{-6}$ however
the spurious velocities observed are nearly 100 times greater.  Further, the
flow field for the standard model includes considerable vorticity and 
significant flows far from the interface itself.  This suggests that 
discretization errors are driving an unstable (and possibly unphysical)
mode of the standard system leading to much larger spurious velocities.
In our model, this mode is absent and discretization errors are not
exaggerated.

One possible origin for unphysical modes in Fig. \ref{spurvel}(a)
and similar models is related to the time scales used.    In
situations where the model parameters have not been measured so that the
length and time scales are uncertain, the viscosity and diffusion constant
may be set for convenience.  Normally this would not affect equilibrium
properties, but if unphysical choices are made, unstable modes like the ones 
being driven by the spurious velocities may be set up.  For instance, unstable
modes may appear if the diffusion constant is set so large that material can 
diffuse faster than it can flow.  To ensure that this is not the case in our
model, we have used a viscosity and diffusion constant that
were measured in molecular dynamics simulations for
similar fluids, $\eta \approx 3 \,\epsilon \tau/\sigma$ \cite{TR90} and 
$D=0.1 \,\sigma^2/\tau$ \cite{DR01}.

\section{Summary and Conclusions}

This paper presents detailed molecular dynamics simulation results
for a binary mixture of simple fluids and uses them to construct
a square gradient theory that can be used for realistic mesoscale
modeling.
The MD simulations examine the linear response, interfacial tension,
and interfacial width as a function of density, temperature, and
the repulsion between different species.
Two remarkable conclusions arise from the linear response results
(Sec. \ref{linresponse}).
The first is that a square gradient theory is capable of
quantitatively describing the response down to wavelengths
that are comparable to the molecular spacing ($< 2\sigma$).
This implies that mesoscale models may have a much wider
range of applicability than might be expected.
The second surprising result is that the elastic constant
for density changes, $K_\rho$, is {\it negative}.
As a result, the system is more susceptible to fluctuations
at shorter wavelengths.
Indeed, it is only stabilized at small scales by atomic discreteness
\cite{footnote}.
This prevents density fluctuations on length scales less than $\sigma$ where
the total response coefficient $L_{\rho \rho}$ becomes negative.

Studies of interfacial properties (Sec. \ref{MDinterface}) show that
the common assumption of incompressibility is not valid.
This is related to the observation that $K_\rho <0$, which
lowers the free energy cost of localized density fluctuations.
Although the density change is small, it can reduce the interfacial
tension by a factor of 2--4.

The density, order parameter and surface stress were evaluated as
a function of position normal to the interface, and
used to determine interfacial widths.
The variation in width with system size is consistent with
broadening by thermal capillary waves. 
Comparing the scaling of widths from the stress and order
parameter, allows all the parameters of the capillary model
to be determined independently.

Fits to linear response about states on the coexistence curve showed
an impressive ability to predict interfacial properties.
Predicted values of the surface tension (Fig. \ref{stension})
and the density deficit at the interface (Sec. \ref{LBresults})
are nearly within the statistical error bars of the MD results.
This is particularly surprising given the large change in order
parameter through the interface and small interface width
(Fig. \ref{intwidth}).
The main discrepancy between the MD results and mesoscale
simulations is that the latter do not include interface
broadening by capillary waves.

Many LB models have been found to produce spurious velocities
around curved interfaces.
While some discretization error is expected, it appears that
these errors are amplified when the time scales in the
LB model are chosen arbitrarily or for computational convenience.
Using time scales derived from MD simulations prevents unphysical
choices that, for example, allow material to diffuse more rapidly
than it flows.
Fig. \ref{spurvel} shows that using MD parameters can
reduce spurious velocities by around three orders of magnitude.

The final square gradient theory (Table \ref{psitable}) has a
simple analytic form and provides
an excellent fit to all MD results for phase coexistence, linear response,
and interfacial properties over a wide range of densities.
This is particularly important for future studies of nonequilibrium
phenomena such as pinchoff or contact line motion.
Dynamic processes will lead to variations in local density and
order parameter that will in turn lead to variations in
local interfacial stress.
These variations will have important implications for the
dynamics, producing Marangoni-like effects or changing the
wavelength dependent response of the interface.
Our results for the influence of density changes on interfacial
tension indicate that these effects may be quite large.
While simpler theories (e.g. Appendix \ref{psivar}) may be
able to describe equilibrium configurations, they do
not include these important dependences on local density.
More complex theories that are not guided by MD results
are unlikely to include important effects such as a
negative $K_{\rho}$.
It will be interesting to explore the dynamic consequences of
such effects in future work.

\begin{acknowledgements}
This material is based upon work supported by the National Science
Foundation under Grant No. 0083286.  The molecular 
dynamics simulations were performed using a variant of LAMMPS from 
Sandia National Laboratories \cite{lammps}.
\end{acknowledgements}

\appendix

\section{Elastic Constant Variations}
\label{Kvar}

Frequently the free energy is parameterized as a function of the individual
species densities $\rho_1$ and $\rho_2$. The free energy can then be 
expressed as
\begin{eqnarray}
{\cal F}&=& k_B T \int d{\bf r}\, \left\{
  \psi+\frac{1}{2}  K_{11}(\nabla \rho_1)^2+\right. \nonumber\\
& & \qquad \qquad \quad \left.
\frac{1}{2} K_{22} (\nabla
  \rho_2)^2 +K_{12}\nabla \rho_1 \cdot \nabla \rho_2\right\}.
\label{freegen}
\end{eqnarray}
The resulting elastic constants are linearly related to those in 
Eq.(\ref{free}): $K_{11}=K_\rho+K_\Phi+2 K_{\rho\Phi}$, 
$K_{11}=K_\rho+K_\Phi-2 K_{\rho\Phi}$, and $K_{12}=K_\rho-K_\Phi$.  

Another order parameter that is often used, the relative concentration, 
is $\phi=\Phi/\rho$.  The free energy becomes
\begin{eqnarray}
{\cal F}&=& k_B T \int d{\bf r} \left\{\psi+
\frac{1}{2}  k_\rho (\nabla \rho)^2+ \right. \nonumber\\
& & \qquad \qquad \quad \left. \frac{1}{2} k_\phi (\nabla \phi)^2+ k_{\rho\phi} \nabla \rho \cdot \nabla \phi
  \right\}.
\label{free2}
\end{eqnarray}
These elastic constants have a more complex mapping to those in the main text,
\begin{eqnarray}
K_\rho &=& k_\rho+k_\phi \Phi^2/\rho^4-2 k_{\rho\phi}\Phi/\rho^2,  \\
K_\Phi &=& k_\phi/\rho^2, \\
K_{\rho\Phi} &=& k_{\rho\phi}/\rho-k_\phi \Phi/\rho^3.
\end{eqnarray}
It is obvious from these relations that the $K$'s and $k$'s can not both
be independent of $\rho$ or $\Phi$.  
There can be advantages to using either $\phi$ or $\Phi$ in different 
physical situations.  It turns out that for linear response measurements and
for the lattice Boltzmann algorithm it is convenient to work with $\Phi$.

As the elastic constants can vary as a function of $\Phi$ and $\rho$ this
should, in principle, be taken into account in Eq.(\ref{muPhi}).  For the 
elastic constants used in the main text, the full Euler-Lagrange equations are
\begin{eqnarray}
\mu_\rho&=& \frac{\partial \psi}{\partial
  \rho}-K_\rho \nabla^2 \rho-K_{\rho\Phi}\nabla^2 \Phi- 
\frac{1}{2}\frac{\partial K_\rho}{\partial\rho}(\nabla \rho)^2+\nonumber\\
& &
\left[\frac{1}{2}\frac{\partial K_\Phi}{\partial\rho}-\frac{\partial K_{\rho\Phi}}{\partial \Phi}\right](\nabla \Phi)^2-
\frac{\partial K_{\rho}}{\partial \Phi}\nabla \rho\cdot\nabla\Phi,
\label{fullmurho}\\
\mu_\Phi&=& \frac{\partial \psi}{\partial
  \Phi}-K_\Phi \nabla^2 \Phi-K_{\rho\Phi}\nabla^2 \rho-\frac{1}{2}\frac{\partial K_\Phi}{\partial\Phi}(\nabla \Phi)^2+\nonumber\\
& & \left[\frac{1}{2}\frac{\partial K_\rho}{\partial\Phi}-\frac{\partial K_{\rho\Phi}}{\partial \rho}\right](\nabla \rho)^2-
\frac{\partial K_\Phi}{\partial\rho}\nabla \Phi \cdot \nabla \rho.
\label{fullmuphi}
\end{eqnarray}
Clearly it would be desirable if some of these terms, especially those
involving variations of the elastic constants, were negligible.  For the
linear response regime it is straightforward to show that these additional
terms are order  $\delta^2$ (cf. Equation (\ref{linreseqn})).  However,
we must still investigate their importance for interfaces.

\begin{figure}
\includegraphics[width=3.2in]{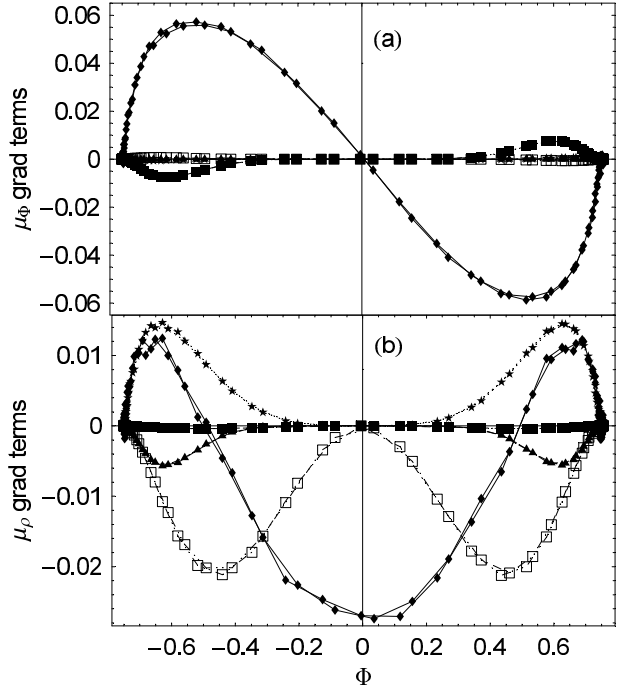}
\caption{The individual gradient term contributions to the potentials 
(a) $\mu_\Phi$ (Eq.(\ref{fullmuphi})) 
and (b) $\mu_\rho$ (Eq.(\ref{fullmurho})) through two interfaces
in a system at average density $\rho=0.85$ with $\epsilon^*=6$.  
For $\mu_\Phi$, the term $K_\Phi \nabla^2 \Phi$ ($\blacklozenge$) dominates,
$\frac{1}{2}(\partial K_\Phi/\partial \Phi)(\nabla \Phi)^2$ ($\blacksquare$) 
is small but visibly different from zero, and 
$\frac{1}{2}(\partial K_\rho/\partial\Phi)(\nabla \rho)^2$ ($\star$), 
$(\partial K_{\rho\Phi}/\partial \rho)(\nabla \rho)^2$ ($\Box$), and 
$(\partial K_\Phi/\partial\rho)\nabla\Phi \cdot \nabla\rho$ ($\blacktriangle$) 
are effectively zero.
(b) For $\mu_\rho$ the significant terms are $K_\rho \nabla^2 \rho$
($\blacklozenge$), $K_{\rho\Phi} \nabla^2 \Phi$ ($\star$) and
$(\partial K_{\rho\Phi}/\partial \Phi)(\nabla \Phi)^2$ ($\Box$).
The $\frac{1}{2}(\partial K_\Phi/\partial \rho)(\nabla \Phi)^2$ 
($\blacktriangle$) term is small but visibly different from zero.  
The other terms, 
$\frac{1}{2}(\partial K_\rho/\partial\rho)(\nabla \rho)^2$ ($\blacksquare$)
and $(\partial K_{\rho}/\partial \Phi)\nabla \rho\cdot\nabla\Phi$ ($\triangle$)
are effectively zero.}
\label{mugrads}
\end{figure}

Figure~\ref{mugrads} shows various contributions from gradients
to the potentials of Eq. (\ref{fullmurho}) and (\ref{fullmuphi}) in a typical 
interface.
As can be seen, the terms that dominate have been kept in Eqs. (\ref{murho}) 
and (\ref{muPhi}).  The remaining terms are quite small, or essentially
zero, thus justifying ignoring them in the main text.  Note that some
terms that are quite small in interfaces, such as $K_{\rho\Phi}\nabla^2 \rho$ 
in $\mu_\Phi$, are required when looking at the linear response where 
$\nabla \rho$ and $\nabla \Phi$ are of comparable magnitude.

\section{Fits to Flory-Huggins Free Energy}
\label{psivar}

There are a number of free energy functionals in common use
to study fluid mixtures.  For polymer mixtures, the Flory-Huggins
free energy density is most commonly used.  For monomers the free energy
density divided by $k_B T$ is \cite{grest}
\begin{equation}
\psi_{FH}=\rho_1 \ln \left(\frac{\rho_1}{\rho}\right)+ \rho_2 \ln \left(\frac{\rho_2}{\rho}\right) + \chi \frac{\rho_1 \rho_2}{\rho}
\label{psiFH}
\end{equation}
where $\chi$ is the only free parameter.
This is just a modified entropy of mixing and, in order to obtain 
the bulk pressure, one must add to this an additional
function $\psi_0$ of $\rho$ alone,
\begin{equation}
\psi=\psi_0 + \psi_{FH}.
\label{fullpsiFH}
\end{equation}
If one were to follow the spirit of the derivation of the Flory-Huggins
model, $\psi_0$ should be determined primarily from the entropy of an
ideal gas plus some quadratic terms to correct for energy interactions.
In practice it is unrealistic to expect such a construction to work.  We
shall use the same form for $\psi_0$ as we used for $A_0$ in 
Eq.(\ref{A0form}).  There is also some ambiguity in the definition of the
$\chi$ term in the Flory-Huggins free energy.  Some authors use a slightly
different term, $\chi \rho_1 \rho_2/(v \rho^2)$ where $v$ is a reference 
volume \cite{Larson}.  If we allow $\chi$ to be a function of $\rho$ then
both terms are equivalent but the meaning of $\chi$ will be slightly
different.  

\begin{figure}
\includegraphics[width=3.2in]{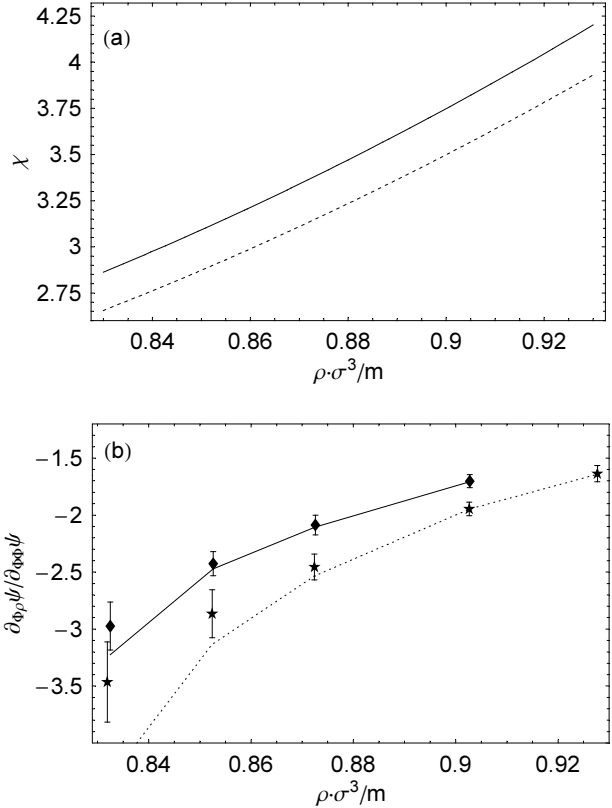}
\caption{$\chi$ parameter (a) obtained from the fits to $\Phi_{co}(\rho)$ 
(Fig.~\ref{rhophi}) and $d\Phi_{co}/d\rho$ (b).  Solid and dotted lines 
correspond to fits for $\epsilon^*=6$ and $5$ respectively.  Molecular dynamics
data in (b) are for $\epsilon^*=5$ ($\star$) and $6$ ($\blacklozenge$).}
\label{chi}
\end{figure}

We obtain $\chi$ by fitting $\Phi_{co}$ and $d\Phi_{co}/d\rho$ as a 
function of $\rho$.  In the bulk states $\mu_\Phi$ is zero and there are 
no gradients so that Eq.(\ref{muPhi}) requires that on the coexistence line
\begin{equation}
0=\frac{\partial \psi}{\partial \Phi}=\frac{\partial \psi_{FH}}{\partial \Phi}=
\frac{1}{2}\ln\left[\frac{1+\Phi/\rho}{1-\Phi/\rho} \right]-\frac{1}{2} \chi \frac{\Phi}{\rho}.
\end{equation}
Eq.(\ref{dsofA0}) also holds, and if one evaluates these derivatives for the
Flory-Huggins free energy one can obtain the relation
\begin{eqnarray}
&&\left(\frac{1}{1+\Phi/\rho}+\frac{1}{1-\Phi/\rho}\right)
\left(\frac{\Phi}{\rho}+\frac{d\Phi_{co}}{d \rho}\right)=\qquad\qquad \nonumber\\
&& \qquad\qquad\qquad \chi \left(\frac{\Phi}{\rho}+\frac{d\Phi_{co}}{d \rho}\right) -
\Phi \frac{\partial \chi}{\partial \rho},
\end{eqnarray}
where $d\Phi_{co}/d \rho$ is evaluated using Eq.(\ref{dsofA0}) and the
measured values of the second derivatives.  If we take $\chi$ to be
a quadratic function of density,
\begin{equation}
\chi=\chi_0+\chi_1 \rho+\chi_2 \rho^2,
\label{chiform}
\end{equation}
then we can do a simultaneous fit to these two equations using a 
straightforward weighted linear regression (there should not
be any conflict between them as we are just fitting $\Phi_{co}(\rho)$
and the derivative of this function $d\Phi_{co}/d\rho$).  The weights
are computed from the statistical errors of the measurements of
$\Phi_{co}$ and the second derivatives.  We use standard methods to find 
the statistical errors of the derived quantities.  The resulting fits are 
shown in Fig.~\ref{chi} and the parameters are given in Table~\ref{chitab}.

Next we fit the bulk pressure $p_0$ to obtain $\psi_0$.  Unlike the fits of
the equilibrium $p_0$ in the main text, which involved only $A_0$, there is 
now a contribution from the $\Phi$ dependent terms in the free energy.  In
particular,
\begin{equation}
p_0=\rho \frac{\partial \psi_0}{\partial \rho} - \psi_0+
\frac{1}{4}k_B T (\rho^2-\Phi^2)\frac{\partial \chi}{\partial \rho}.
\end{equation}
One could, in principle, also use the information about $d p_0/d\rho$
that we used in fitting $A_0$ for the single phase free energy.
However, the additional terms make a combined fit extremely messy and of 
dubious value.
One still needs a parameterization of $\psi_0$ in terms of $\rho$ and
we use the same functional form as was used for $A_0$ (although the
numerical values for $a_{00}$ etc. will, or course be different).
The resulting fit to $p_0$ is shown as a dashed line in Fig.~\ref{p0fit}
and the parameters are given in Table~\ref{chitab}.  

As all parameters in the Flory-Huggins free energy are now determined, we
can now compare the quantities not explicitly used in the fits.
 Fig.~\ref{d2fit} shows the second derivatives of the bulk free energy as
measured from linear response in the molecular dynamics simulations and
from the fits.   The Flory-Huggins theory overestimates the concavity 
of the free energy minima.  Fig.~\ref{stension} shows the surface tension 
derived from a lattice Boltzmann implementation of the fit to the
Flory-Huggins theory.  The agreement is remarkably good.  However, 
Flory-Huggins normally assumes incompressiblity.  If we had assumed that
the density, and therefore $\chi$ was constant, the surface tension would be
much too large.  

It is also worthwhile to compare the values of $\chi$ 
obtained here to other methods of estimating $\chi$.   For long polymers
at $\rho=0.85 m/\sigma^3$,
Grest and coworkers \cite{grest} have estimated $\chi$ in two ways.
Using a so-called one-fluid approximation, they estimate 
$\chi \approx 0.76 \epsilon^* \epsilon/k_B T$.
Using an incompressible random phase approximation to evaluate the static
structure factor, they obtain a larger value of 
$\chi \approx 1.0 \epsilon^* \epsilon/k_B T$.
Our results correspond to a somewhat smaller value of about
$0.54 \epsilon^* \epsilon/k_B T$, which is not surprising given
that we consider simple monomers.

\begin{table}
\begin{tabular}{c|c|c}
\hline
\hline  
$\epsilon^*$&                  5                        &            6     \\
\hline
$a_{0}$     & $-31.51 \pm 2.9\, \frac{m}{\sigma \tau^2}$& $-26.50 \pm 6.3\, \frac{m}{\sigma \tau^2}$ \\
$a_{1}$     & $-92.33\pm 6.7\,\frac{\sigma^2}{\tau^2}$ & $-80.18\pm 14\,\frac{\sigma^2}{\tau^2}$ \\
$a_{2}$     & $77.0\pm 3.8\,\frac{\sigma^5}{m \tau^2}$ & $69.7\pm 8.3\,\frac{\sigma^5}{m \tau^2}$ \\
\hline
$\chi_{0}$     & $10.15 \pm 4.7\,$                      & $10.68 \pm 5.0$ \\
$\chi_{1}$     & $-28.47 \pm 10.4\, \frac{\sigma^3}{m}$   & $-29.78 \pm 11.21\,\frac{\sigma^3}{m} $ \\
$\chi_{2}$     & $23.42 \pm 5.7\,  \frac{\sigma^6}{m^2}$   & $24.52 \pm 6.32\, \frac{\sigma^6}{m^2}$ \\
\hline
\hline
\end{tabular} 
\caption{Table of parameters for the Flory-Huggins free energy given in 
Eq.s (\ref{psiFH}) and (\ref{fullpsiFH}).
Parameters are defined in Eqs. (\ref{A0form}) and (\ref{chiform}).  
As for the parameters given in the main text, the data were measured for 
$0.82\, m/\sigma^3 < \rho < 0.925 \,m/\sigma^3$ and care should be taken in 
extrapolating outside of the measured range. 
\label{chitab}}
\end{table}


\begin{thebibliography}{99.}
\bibitem{footnote}
Stability could be enforced by adding a higher order gradient with a positive
coefficient.  However, a very high order may be needed and this would
 complicate any numerical scheme without providing a clear benefit.
\end{thebibliography}
\end{document}